\documentclass[11pt,a4paper,twoside,doc,apacite]{apa6} % doc man

\usepackage{amssymb,amsmath}
\usepackage{apacite}
\usepackage{commath}
\usepackage{array} % to specify the width of cells within the tabular environment
\usepackage{ctable} % for table footnote
\usepackage[american]{babel}
\usepackage{epstopdf} % for .eps files¬ß
\usepackage{eurosym} % ? symbol
\usepackage{float} % for figures
\usepackage{graphicx} % to load graphics 
\usepackage{paralist} % to enumerate sentences/words
\usepackage{setspace} % fine-grained control over line spacing
\usepackage{threeparttable} % for table footnotes
\usepackage{bbm}
\usepackage{pdflscape} % for landscape pages
\usepackage{subfigure} % for two figures in one figure
\usepackage{amsbsy} % for bold symbols
\usepackage{relsize}
\usepackage{color}

\usepackage[toc]{appendix}

\usepackage{listings}
\allowdisplaybreaks[1]

\title{A Tutorial on Bridge Sampling}

\leftheader{}
\shorttitle{A Tutorial on Bridge Sampling}
\author{Quentin F.~Gronau$^1$, Alexandra Sarafoglou$^1$, Dora Matzke$^1$, Alexander Ly$^1$, Udo Boehm$^1$, Maarten Marsman$^1$, David S.~Leslie$^2$, Jonathan J.~Forster$^3$, Eric-Jan Wagenmakers$^1$, Helen Steingroever$^1$}
\affiliation{$^1$ Department of Psychology, University of Amsterdam,\\ The Netherlands \\
                $^2$ Department Mathematics and Statistics, Lancaster University, UK \\
                $^3$ Mathematical Science, University of Southampton, UK \\
\vspace{1cm}
Correspondence concerning this article should be addressed to: Quentin F.~Gronau, Department of Psychology, PO Box 15906, 1001 NK Amsterdam, The Netherlands, E-mail: Quentin.F.Gronau@gmail.com. 
}
\note{}

\abstract{The marginal likelihood plays an important role in many areas of Bayesian statistics such as parameter estimation, model comparison, and model averaging. In most applications, however, the marginal likelihood is not analytically tractable and must be approximated using numerical methods. Here we provide a tutorial on bridge sampling \cite{bennett1976efficient, MengWong1996}, a reliable and relatively straightforward sampling method that allows researchers to obtain the marginal likelihood for models of varying complexity. First, we introduce bridge sampling and three related sampling methods using the beta-binomial model as a running example. We then apply bridge sampling to estimate the marginal likelihood for the Expectancy Valence (EV) model---a popular model for reinforcement learning. Our results indicate that bridge sampling provides accurate estimates for both a single participant and a hierarchical version of the EV model. We conclude that bridge sampling 
is an attractive method for mathematical psychologists who typically aim to approximate the marginal likelihood for a limited set of possibly high-dimensional models.}


\begin{document}

\maketitle

Bayesian statistics has become increasingly popular in mathematical psychology \cite{andrews2013prior, Bayarri201690, poirier2006growth, vanpaemel2016prototypes, verhagen2015evaluating, wetzels2016bayesian}. The Bayesian approach is conceptually simple, theoretically coherent, and easily applied to relatively complex problems. These problems include, for instance, hierarchical modeling (\citeNP{Matzke2015,matzke2009psychological, rouder2005introduction, rouder2005hierarchical, rouder2007signal}) or the comparison of non-nested models \cite{lee2008three, pitt2002toward, shiffrin2008survey}. Three major applications of Bayesian statistics concern parameter estimation, model comparison, and Bayesian model averaging. In all three areas, the marginal likelihood --that is, the probability of the observed data given the model of interest-- plays a central role (see also \citeNP{GelmanMeng1998}).

First, in parameter estimation, we consider a single model and aim to quantify the uncertainty for a parameter of interest $\theta$ after having observed the data $y$. This is realized by means of a posterior distribution that can be obtained using Bayes theorem:

\begin{align} 
  \label{Eq:BayesTheorem}
  p(\theta \mid y) = \cfrac{p(y \mid \theta) \; p(\theta)}{\int p(y \mid \theta^\prime) \; p(\theta^\prime) \, \mathrm{d} \theta^\prime} = \cfrac{\overbrace{p(y \mid \theta)}^{\text{likelihood}} \; \overbrace{p(\theta)}^{\text{prior}}}{\underbrace{p(y)}_{\text{marginal likelihood}}} \; .
\end{align}

\noindent
Here, the marginal likelihood of the data $p(y)$ ensures that the posterior distribution is a proper probability density function (PDF) in the sense that it integrates to 1. This illustrates why in parameter estimation the marginal likelihood is referred to as a normalizing constant.

Second, in model comparison, we consider $m$ ($m \in \mathbb{N}$) competing models, and are interested in the relative plausibility of a particular model $\mathcal{M}_i$ ($i \in \{1, 2, \ldots, m\}$) given the prior model probability and the evidence from the data $y$ (see three special issues on this topic in the \emph{Journal of Mathematical Psychology}: \citeNP{mulder2016editors, myung2000guest, Wagenmakers200699}). This relative plausibility is quantified by the so-called posterior model probability $p(\mathcal{M}_i \mid y)$ of model $\mathcal{M}_i$ given the data $y$ \cite{Berger2005PMP}: 

\begin{equation} \label{E:PMP_ind}
   p(\mathcal{M}_i \mid y) = \frac{p(y \mid \mathcal{M}_i) \; p(\mathcal{M}_i)}{\sum_{j=1}^{m} p(y \mid \mathcal{M}_j ) \; p(\mathcal{M}_j)},
\end{equation}

\noindent where the denominator is the sum of the marginal likelihood times the prior model probability of all $m$ models. In model comparison, the marginal likelihood for a specific model is also referred to as the model evidence \cite{didelot2011likelihood}, the integrated likelihood \cite{kass1995bayes}, the predictive likelihood of the model \cite[chapter 7]{Gamerman2006}, the predictive probability of the data \cite{kass1995bayes}, or the prior predictive density \cite{Ntzoufras2009}. Note that conceptually the marginal likelihood of Equation~\ref{E:PMP_ind} is the same as the marginal likelihood of Equation~\ref{Eq:BayesTheorem}. However, for the latter equation we droped the model index because in parameter estimation we consider only one model.

If only two models $\mathcal{M}_1$ and $\mathcal{M}_2$ are considered, Equation~\ref{E:PMP_ind} can be used to quantify the relative posterior model plausibility of model $\mathcal{M}_1$ compared to model $\mathcal{M}_2$. This relative plausibility is given by the ratio of the posterior probabilities of both models, and is referred to as the posterior model odds:

\begin{equation} \label{E:BF}
\underbrace{\cfrac{p(\mathcal{M}_1\mid y)}{p(\mathcal{M}_2\mid y)}}_{\substack{\text{posterior}\\\text{odds}}} = 
                            \underbrace{\cfrac{p(\mathcal{M}_1)}
                            {p(\mathcal{M}_2)}}_{\substack{\text{prior}\\\text{odds}}} \times
                            \underbrace{\cfrac{p(y\mid \mathcal{M}_1)}
                            {p(y\mid \mathcal{M}_2)}}_{\substack{\text{Bayes}\\\text{factor}}}.
\end{equation}

Equation~\ref{E:BF} illustrates that the
posterior model odds are the product of two factors: The first factor is the ratio of the prior probabilities of both models---the prior model odds. The second factor is the ratio of the marginal likelihoods of both models---the so-called Bayes factor \cite{etzjbs2016, Jeffreys1961BF,  ly2016evaluation, ly2016harold, robert2016expected}. The Bayes factor plays an important role in model comparison and is referred to as the ``standard Bayesian solution to the hypothesis testing and model selection problems'' \cite[p.~648]{lewis1997estimating} and ``the primary tool used in Bayesian inference for hypothesis testing and model selection'' \cite[p.~378]{Berger2006}. 

Third, the marginal likelihood plays an important role in Bayesian model averaging (BMA; \citeNP{hoeting1999bayesian}) where aspects of parameter estimation and model comparison are combined. As in model comparison, BMA considers several models; however, it does not aim to identify a single best model. Instead  it fully acknowledges model uncertainty.  Model averaged parameter inference can be obtained by combining, across all models, the posterior distribution of the parameter of interest weighted by  each model's  posterior model probability, and as such depends on the marginal likelihood of the models. This procedure assumes that the parameter of interest has identical interpretation across the different models. Model averaged predictions can be  obtained in a similar manner.

A problem that arises in all three areas---parameter estimation, model comparison, and BMA---is that an analytical expression of the marginal likelihood can be obtained only for certain restricted examples. This is a pressing problem in Bayesian modeling, and in particular in mathematical psychology where models can be non-linear and equipped with a large number of parameters, especially when the models are implemented in a hierarchical framework. Such a framework incorporates both commonalities and differences between participants of one group by assuming that the model parameters of each participant are drawn from a group-level distribution (for advantages of the Bayesian hierarchical framework see \citeNP{Ahn20114,  navarro2006modeling, rouder2005introduction, rouder2005hierarchical, rouder2008hierarchical, Scheibehenne2015, shiffrin2008survey, Wetzels201014}). For instance, consider a four-parameter Bayesian hierarchical model with four group-level distributions each characterized by two parameters and a group size of 30 participants; this then results in $30 \times 4$ individual-level parameters and $2 \times 4$ group-level parameters for a total of 128 parameters. In sum, even simple models quickly become complex once hierarchical aspects are introduced and this frustrates the derivation of the marginal likelihood. 

To overcome this problem, several Monte Carlo sampling methods have been proposed to approximate the marginal likelihood. In this tutorial we focus on four such methods: the bridge sampling estimator (\citeNP{bennett1976efficient}, Chapter 5 of \citeNP{chen2012monte}, \citeNP{MengWong1996}), and its three commonly used special cases---the naive Monte Carlo estimator, the importance sampling estimator, and the generalized harmonic mean estimator (for alternative methods see \citeNP[Chapter 7]{Gamerman2006}; and for alternative approximation methods relevant to model comparison and BMA see \citeNP{carlin1995bayesian, green1995reversible}).\footnote{The appendix provides a derivation showing that the first three estimators are indeed special cases of the bridge sampler.} As we will illustrate throughout this tutorial,
the bridge sampler is accurate, efficient, and relatively straightforward to implement (e.g., \citeNP{Dicicciokass1997, fruhwirth2004estimating, MengWong1996}). 

The goal of this tutorial is to bring the bridge sampling estimator to the attention of mathematical psychologists. We aim to explain this estimator and facilitate its application by suggesting a step-by-step implementation scheme. To this end, we first show how bridge sampling and the three special cases can be used to approximate the marginal likelihood in a simple beta-binomial model. We begin with the naive Monte Carlo estimator and progressively work our way up---via the importance sampling estimator and the generalized harmonic mean estimator---to the most general case considered: the bridge sampling estimator. This order was chosen such that key concepts are introduced gradually and estimators are of increasing complexity and sophistication. The first three estimators are included in this tutorial with the sole purpose of facilitating the reader's understanding of bridge sampling. In the second part of this tutorial, we outline how the bridge sampling estimator can be used to derive the marginal likelihood for the Expectancy Valence (EV; \citeNP{Busemeyer2002253}) model---a popular, yet relatively complex reinforcement-learning model for the Iowa gambling task \cite{Bechara19947}. We apply bridge sampling to both an individual-level and a hierarchical implementation of the EV model.

Throughout the article, we use the software package R to implement the bridge sampling estimator for the various models. The interested reader is invited to reproduce our results by downloading the code and all relevant materials from our Open Science Framework folder at \url{osf.io/f9cq4}.

\section{Four Sampling Methods to Approximate the Marginal Likelihood}

In this section we outline four standard methods to approximate the marginal likelihood. For more detailed explanations and derivations, we recommend \citeA[Chapter 11]{Ntzoufras2009} and \citeA[Chapter 7]{Gamerman2006}; a comparative review of the different sampling methods is presented in \citeA{Dicicciokass1997}. The marginal likelihood is the probability of the observed data $y$ given a specific model of interest $\mathcal{M}$, and is defined as the integral of the likelihood over the prior:

\begin{align} 
  \label{Eq:ML}
        \underbrace{p(y\mid \mathcal{M})}_{\substack{\text{marginal}\\\text{likelihood}}} = 
           \int \underbrace{p(y\mid \theta, \mathcal{M})}_{\text{likelihood}} \;
                \underbrace{p(\theta\mid \mathcal{M})}_{\text{prior}} \; \mathrm{d}\theta,
\end{align}

\noindent
with $\theta$ a vector containing the model parameters. Equation~\ref{Eq:ML} illustrates that the marginal likelihood can be interpreted as a weighted average of the likelihood of the data given a specific value for $\theta$ where the weight is the a priori plausibility of that specific value. Equation~\ref{Eq:ML} can therefore be written as an expected value:

$$
p(y\mid \mathcal{M}) =  \mathbb{E}_{\text{prior}}\left[p(y \mid \theta, \mathcal{M})\right], 
$$

\noindent
where the expectation is taken with respect to the prior distribution. This idea is central to the four sampling methods that we discuss in this tutorial. 

\subsection{Introduction of the Running Example: The Beta-Binomial Model}

Our running example focuses on estimating the marginal likelihood for a binomial model assuming a uniform prior on the rate parameter $\theta$ (i.e., the beta-binomial model). Consider a single participant who answered $k = 2$ out of $n = 10$ true/false questions correctly. Assume that the number of correct answers follows a binomial distribution, that is, $k \ \sim \text{Binomial}(n, \theta)$ with $\theta \in (0, 1)$, where $\theta$ represents the latent probability for answering any one question correctly. The probability mass function (PMF) of the binomial distribution is given by:

\begin{equation}
  \label{Eq:BinomialPMF}
  \text{Binomial}(k \mid n, \theta) = \binom{n}{k} \theta^k (1 - \theta)^{n - k},
\end{equation}

\noindent
where $k, n \in \mathbb{Z}_{\geq 0}$, and $k \leq n$. The PMF of the binomial distribution serves as the likelihood function in our running example.

\begin{figure}
	\centering
    \includegraphics[width=0.8\textwidth]{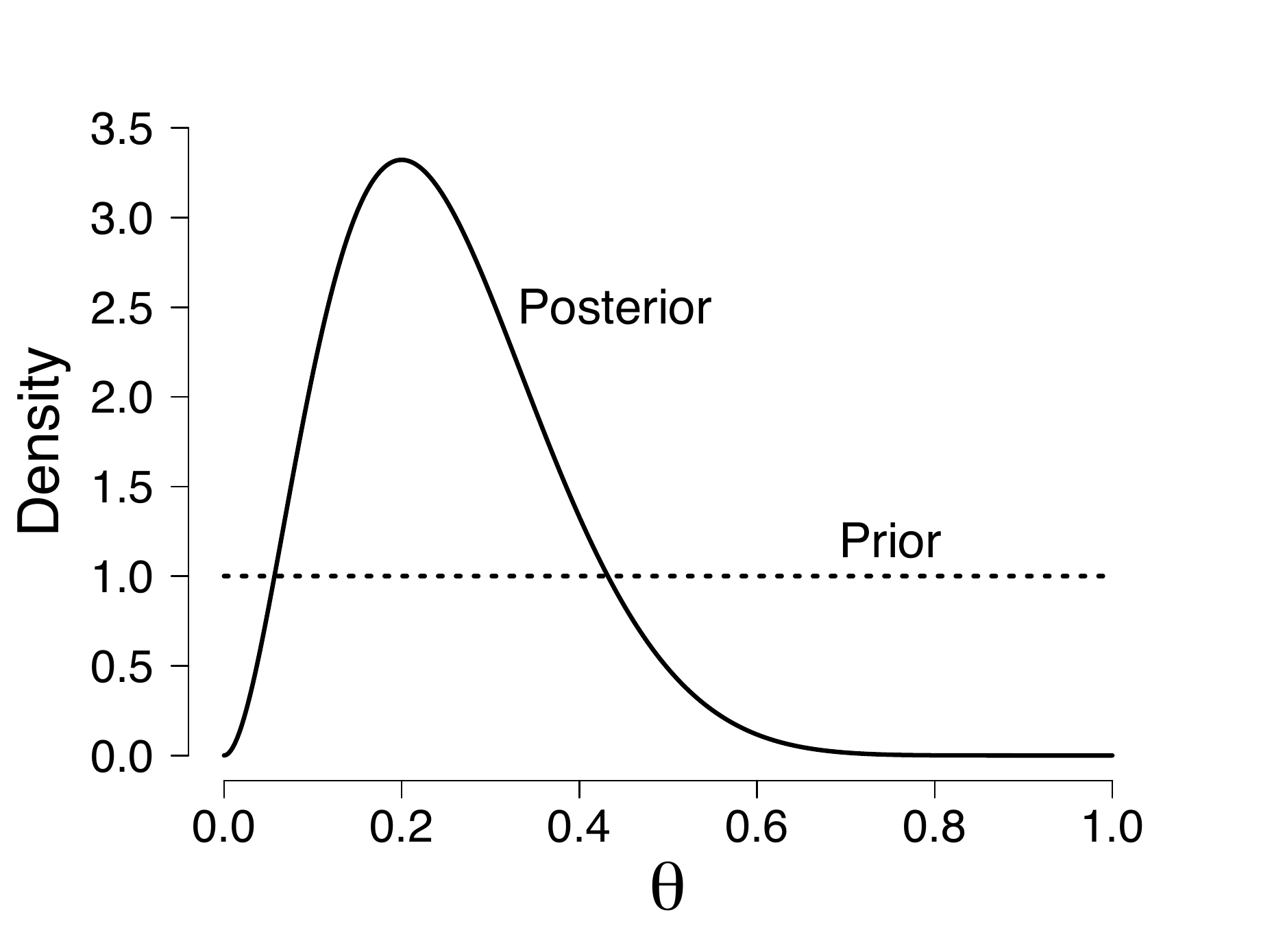}
	\caption{Prior and posterior distribution for the  rate parameter $\theta$ from the beta-binomial model. The $\text{Beta}(1, 1)$ prior on the rate parameter $\theta$ is represented by the dotted line; the $\text{Beta}(3, 9)$ posterior distribution is represented by the solid line and was obtained after having observed 2 correct responses out of 10 trials. Available at \url{https://tinyurl.com/yc8bw98v} under CC license \url{https://creativecommons.org/licenses/by/2.0/}.}
	\label{F:BinomialPriorPost}
\end{figure}

In the Bayesian framework, we also have to specify the prior distribution of the model parameters; the prior distribution expresses our knowledge about the parameters before the data have been observed. In our running example, we assume that all values of $\theta$ are equally likely a priori. This prior belief is captured by a uniform distribution across the range of $\theta$, that is, $\theta \sim \text{Uniform}(0, 1)$ which can equivalently be written in terms of a beta distribution $\theta \sim \text{Beta}(1, 1)$. This prior distribution is represented by the dotted line in Figure~\ref{F:BinomialPriorPost}. It is evident that the density of the prior distribution equals 1 for all values of $\theta$. One advantage of expressing the prior distribution by a beta distribution is that its two parameters (i.e., in its general form the shape parameters $\alpha$ and $\beta$) can be thought of as counts of ``prior successes'' and ``prior failures'', respectively. In its general form, the PDF of a $\text{Beta}(\alpha, \beta)$ distribution ($\alpha, \beta > 0$) is given by:

\begin{equation*}
  \label{Eq:BetaPDF}
  \text{Beta}(\theta; \; \alpha, \beta) = \cfrac{\theta^{\alpha-1} (1 - \theta)^{\beta - 1}}{B(\alpha, \beta)},
\end{equation*}

\noindent
where $B(\alpha, \beta)$ is the beta function that is defined as: $B(\alpha, \beta) = \int_0^1 t^{\alpha - 1}(1 - t)^{\beta-1} \mathrm{d}t = \frac{\Gamma(\alpha) \Gamma(\beta)}{\Gamma(\alpha + \beta)}$, and $\Gamma(n) = (n - 1)!$ for $n \in \mathbb{N}$. 

\subsubsection{Analytical derivation of the marginal likelihood}
As we will see in this section, the beta-binomial model constitutes one of the rare examples where the marginal likelihood is analytic. Assuming a general $k$ and $n$, we obtain the marginal likelihood as:

 \begin{align}
 \label{Eq:AnaSol}
       \nonumber p(k \mid n) &\stackrel{Eq.~\ref{Eq:ML}}{=} \int_0^1 p( k \mid n, \theta) \;
                p(\theta) \, \mathrm{d}\theta 
                = \int_0^1  \binom{n}{k} \theta^k (1 - \theta)^{n - k} \; 1 \; \mathrm{d}\theta \\  
                & \;\; = \binom{n}{k} B(k+1, n-k+1) 
               = \cfrac{1}{n + 1} \; , 
\end{align}

\noindent
where we suppress the ``model'' in the conditioning part of the probability statements because we focus on a single model in this running example. Using $k=2$ and $n=10$ of our example, we obtain: $p(k = 2 \mid n = 10) = 1 / 11 \approx 0.0909$. This value will be estimated in the remainder of the running example using  the naive Monte Carlo estimator,  the importance sampling estimator, the generalized harmonic  mean estimator, and finally the bridge sampling estimator. 

As we will see below, the importance sampling, generalized harmonic mean estimator, and bridge sampling estimator require samples from the posterior distribution. These samples can be obtained  using computer software such as WinBUGS \cite{lunn2000winbugs}, JAGS \cite{Plummer2003}, or Stan \cite{rstan}, even when the marginal likelihood that functions here as a normalizing constant is not known (Equation~\ref{Eq:BayesTheorem}). However,  in our running example MCMC samples are not required because we can derive an analytical expression of the posterior distribution for $\theta$ after having observed the data. Using the analytic expression of the marginal likelihood (Equation~\ref{Eq:AnaSol}) and Bayes theorem, we obtain:

\begin{align*}
p(\theta \mid k, n) &= \cfrac{p(k \mid n, \theta) \; p(\theta)}{p(k \mid n)}
  = \cfrac{\binom{n}{k} \theta^k (1 - \theta)^{n - k} \; 1}{\binom{n}{k} B(k+1, n - k + 1)}
  = \cfrac{\theta^k (1 - \theta)^{n - k} }{ B(k+1, n-k+1)} \; ,\\  
\end{align*}
 
 \noindent
which we recognize as the PDF of the $\text{Beta}(k+1, n-k+1)$ distribution. Thus, if we assume a uniform prior on $\theta$ and observe $k=2$ correct responses out of $n=10$ trials, we obtain a $\text{Beta}(3, 9)$ distribution as posterior distribution. This distribution is represented by the solid line in Figure~\ref{F:BinomialPriorPost}. In general, if $k \mid n, \theta \sim \text{Binomial}(n, \theta)$ and $\theta \sim \text{Beta}(1, 1)$, then $ \theta \mid n, k \sim \text{Beta}(k+1, n-k+1)$.
 
\subsection{Method 1: The Naive Monte Carlo Estimator of the Marginal Likelihood}

The simplest method to approximate the marginal likelihood is provided by the naive Monte Carlo estimator \cite{hammersley1964monte, Raftery1991}. This method uses the standard definition of the marginal likelihood (Equation~\ref{Eq:ML}), and relies on the central idea that the marginal likelihood can be written as an expected value with respect to the prior distribution, that is, $p(y) = \mathbb{E}_\text{prior} \left[p(y\mid \theta)\right]$. This expected value of the likelihood of the data with respect to the prior can be approximated by evaluating the likelihood in $N$ samples from the prior distribution for $\theta$  and averaging the resulting values. This yields the naive Monte Carlo estimator $\hat p_1(y)$:

\begin{align} \label{Eq:NMCE}
        \hat p_1 (y) = 
        \underbrace{\cfrac{1}{N} \sum_{i = 1}^N p(y\mid \tilde \theta_i)}_{\text{average likelihood}}, \; \;
         \underbrace{\tilde \theta_i \sim p(\theta)}_{\substack{\text{samples from the}\\ \text{prior distribution}}}.
\end{align}

\subsubsection{Running example}

\begin{figure}[!bt]
	\centering
    \includegraphics[width=0.8\textwidth]{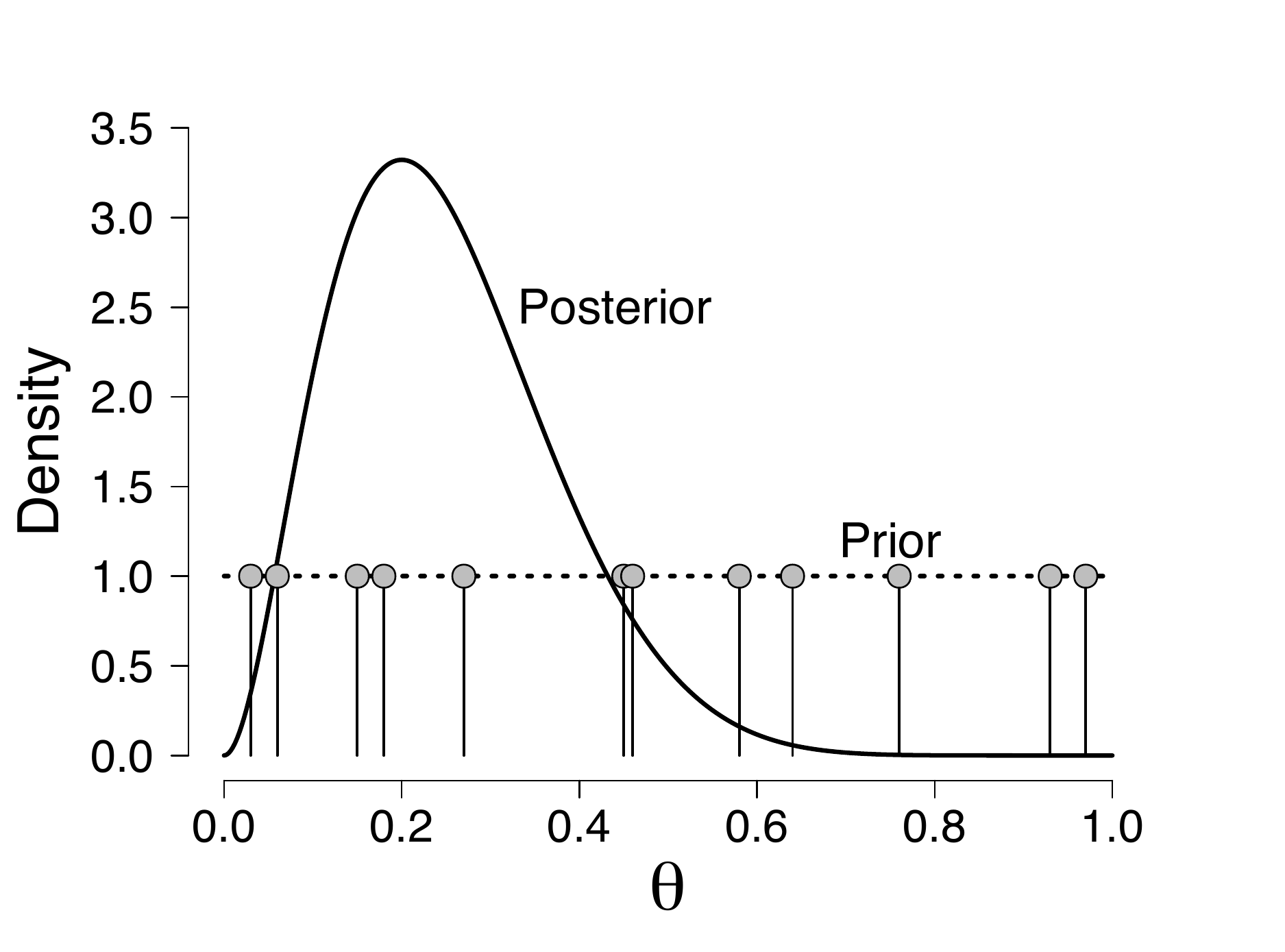}
	\caption{Illustration of the naive Monte Carlo estimator for the beta-binomial example. The dotted line represents the prior distribution and the solid line represents the posterior distribution that was obtained after having observed 2 correct responses out of 10 trials. The gray dots represent the 12 samples $\{ \tilde \theta_1, \tilde \theta_2, \ldots, \tilde \theta_{12} \}$ randomly drawn from the $\text{Beta}(1, 1)$ prior distribution. Available at \url{https://tinyurl.com/y8uf6t8f} under CC license \url{https://creativecommons.org/licenses/by/2.0/}.}
	\label{PriorSample}
\end{figure}

To obtain the naive Monte Carlo estimate of the marginal likelihood in our running example, we need $N$ samples from the $\text{Beta}(1, 1)$ prior distribution for $\theta$. For illustrative purposes, we limit the number of samples to 12 whereas in practice one should take $N$ to be very large. We obtain the following samples:

\begin{align*}
\{ \tilde \theta_1, \tilde \theta_2, \ldots, \tilde \theta_{12} \} =& \{0.58, 0.76, 0.03, 0.93, 0.27, 0.97, 0.45, 0.46, 0.18, 0.64, 0.06, 0.15\},
\end{align*}

\noindent
where we use the tilde symbol to emphasize that we refer to a sampled value. All sampled values are represented by the gray dots in Figure \ref{PriorSample}.

Following Equation~\ref{Eq:NMCE}, the next step is to calculate the likelihood (Equation~\ref{Eq:BinomialPMF}) for each $\tilde \theta_i$, and then to average all obtained likelihood values. This yields the naive Monte Carlo estimate of the marginal likelihood:

\begin{align*}
\hat p_1 (k = 2 \mid n = 10) &=  \cfrac{1}{12} \sum_{i = 1}^{12} p(k = 2 \mid
        n = 10, \tilde \theta_i) 
        =  \cfrac{1}{12} \sum_{i = 1}^{12} \binom{n}{k} (\tilde \theta_i)^k (1 - \tilde \theta_i)^{n - k} \\
               &=  \cfrac{1}{12} \binom{10}{2}
        \left(0.58^2 (1 - 0.58)^8 + \ldots + 0.15^2 (1 - 0.15)^8 \right) \\
               & =  0.0945.
\end{align*}

\subsection{Method 2: The Importance Sampling Estimator of the Marginal Likelihood}

The naive Monte Carlo estimator introduced in the last section performs well if the prior and posterior distribution have a similar shape and strong overlap. However, the estimator is unstable if the posterior distribution is peaked relative to the prior \cite<e.g.,>{Gamerman2006, Ntzoufras2009}. In such a situation, most of the sampled values for $\theta$ result in likelihood values close to zero and contribute only minimally to the estimate. This means that those few samples that result in high likelihood values dominate estimates of the marginal likelihood. Consequently, the variance of the estimator is increased \cite{NewtonRaftery1996, Pajor2016}.\footnote{The interested reader is referred to \citeA{Pajor2016} for a recent improvement on the calculation of the naive Monte Carlo estimator. The proposed improvement involves trimming the prior distribution in such a way that regions with low likelihood values are eliminated, thereby increasing the accuracy and efficiency of the estimator.}

The importance sampling estimator, on the other hand, overcomes this shortcoming by boosting sampled values in regions of the parameter space where the integrand of Equation~\ref{Eq:ML} is large. This is realized by using samples from a so-called importance density $g_{IS}(\theta)$ instead of the prior distribution. 
The advantage of sampling from an importance density is that values for $\theta$ that result in high likelihood values are sampled most frequently, whereas values for $\theta$ with low likelihood values are sampled only rarely. \par
To derive the importance sampling estimator, Equation~\ref{Eq:ML} is used as starting point which is then extended by the importance density $g_{IS}(\theta)$:

\begin{align*}
        p(y) &= \int p(y\mid\theta) \; p(\theta) \; \mathrm{d}\theta
                         = \int 
       p(y\mid \theta) \;
              p(\theta)
               \; 
             \cfrac{g_{IS}(\theta)}{g_{IS}(\theta)} \; \mathrm{d}\theta 
                         = \int 
        \cfrac{{p(y\mid \theta)} \;
              p(\theta)}
              {g_{IS}(\theta)} \; 
             g_{IS}(\theta) \; \mathrm{d}\theta \\
&= \mathbb{E}_{g_{IS}(\theta)}\left(\cfrac{p(y\mid\theta) \;
               p(\theta)}{g_{IS}(\theta)}\right).               
\end{align*}

\noindent This yields the importance sampling estimator $\hat p_2(y)$:

\begin{align} \label{Eq:ISE}
        \hat p_2(y) = \underbrace{\cfrac{1}{N} \sum_{i = 1}^N
        \cfrac{p(y\mid \tilde \theta_i) \; p(\tilde \theta_i)}
             {g_{IS}(\tilde \theta_i)}}_{\text{average adjusted likelihood}}, \; \;
             \underbrace{\tilde \theta_i \sim g_{IS}(\theta).}_{\substack{\text{samples from the}\\ \text{importance density}}}
\end{align}

A suitable importance density should (1) be easy to evaluate; (2) have the same domain as the posterior distribution; (3) closely resemble the posterior distribution; and (4) have fatter tails than the posterior distribution \cite{Neal2001, VanderkerckhoveIS}. The latter criterion ensures that values in the tails of the distribution cannot misleadingly dominate the estimate  \cite{Neal2001}.\footnote{To illustrate the need for an importance density with fatter tails than the posterior distribution, imagine you sample from the tail region of an importance density with thinner tails. In this case, the numerator in Equation~\ref{Eq:ISE} would be substantially larger than the denominator resulting in a very large ratio. Since this specific ratio is only one component of the sum displayed in Equation~\ref{Eq:ISE}, this component would dominate the importance sampling estimate. Hence, thinner tails of the importance density run the risk of producing unstable estimates across repeated computations. In fact, the estimator may have infinite variance (e.g., \citeNP{ionides2008truncated, OwenZhou2000}).} 

\subsubsection{Running example}

To obtain the importance sampling estimate of the marginal likelihood in our running example, we first need to choose an importance density $g_{IS}(\theta)$. An importance density that fulfills the four above mentioned desiderata is a mixture between a beta density that provides the best fit to the posterior distribution and a uniform density across the range of $\theta$ \cite{VanderkerckhoveIS}. The relative impact of the uniform density is quantified by a mixture weight $\gamma$ that ranges between $0$ and $1$. The larger $\gamma$, the higher the influence of the uniform density resulting in a less peaked distribution with thick tails. If $\gamma = 1$, the beta mixture density simplifies to the uniform distribution on $[0, 1]$;\footnote{In our running example, the importance sampling estimator then reduces to the naive Monte Carlo estimator.} and if $\gamma = 0$, the beta mixture density simplifies to the beta density that provides the best fit to the posterior distribution.  

In our specific example, we already know that the $\text{Beta}(3, 9)$ density is the beta density that provides the best fit to the posterior distribution because this is the analytic expression of the posterior distribution. However, to demonstrate the general case, we show how we can find the beta distribution with the best fit to the posterior distribution using the method of moments. This particular method works as follows. First, we draw samples from our $\text{Beta}(3, 9)$ posterior distribution and obtain:\footnote{Note that, when the analytical expression of the posterior distribution is not known, posterior samples can be obtained using computer software such as WinBUGS, JAGS, or Stan, even when the marginal likelihood that functions here as a normalizing constant is not known (Equation~\ref{Eq:BayesTheorem}).}

\begin{align*}
\{ \theta^*_1, \theta^*_2, \ldots, \theta^*_{12} \} =& \{0.22, 0.16, 0.09, 0.35, 0.06, 0.27, 0.26, 0.41, 0.20, 0.43, 0.21, 0.12\}.
\end{align*}

\noindent
Note that here we use $\theta_i^*$ to refer to the $i^{\text{th}}$ sample from the posterior distribution to distinguish it from the previously used $\tilde \theta_i$---the $i^{\text{th}}$ sample from a distribution other than the posterior distribution, such as a prior distribution or an importance density. Second, we compute the mean and variance of these posterior samples. We obtain a mean of $\bar \theta^* = 0.232$ and a variance of $s_{\theta*}^2 = 0.014$.  

Third, knowing that, if $X \sim \text{Beta}(\alpha, \beta)$, then $\mathbb{E}(X) = \alpha / (\alpha + \beta)$ and $V(X) = \alpha \beta / \left [ (\alpha + \beta)^2 (\alpha + \beta + 1) \right ]$, we obtain the following method of moment estimates for $\alpha$ and $\beta$:

\begin{align*}
        \hat \alpha &= \bar \theta^* \left( \cfrac{\bar \theta^* (1 - \bar
        \theta^*)}{s_{\theta^*}^2} - 1 \right) 
  = 0.232 \left( \cfrac{0.232(1 - 0.232)}{0.014} - 1 \right)
  = 2.721,
\end{align*}

\begin{align*}
        \hat \beta &= (1 - \bar \theta^*) \left(\cfrac{\bar \theta^* (1-\bar\theta^*)}
              {s_{\theta^*}^2} - 1\right)
  = (1-0.232) \left( \cfrac{0.232(1 - 0.232)}{0.014} - 1 \right)
  = 9.006.
\end{align*}

\begin{figure}[!bt]
	\centering
    \includegraphics[width=0.8\textwidth]{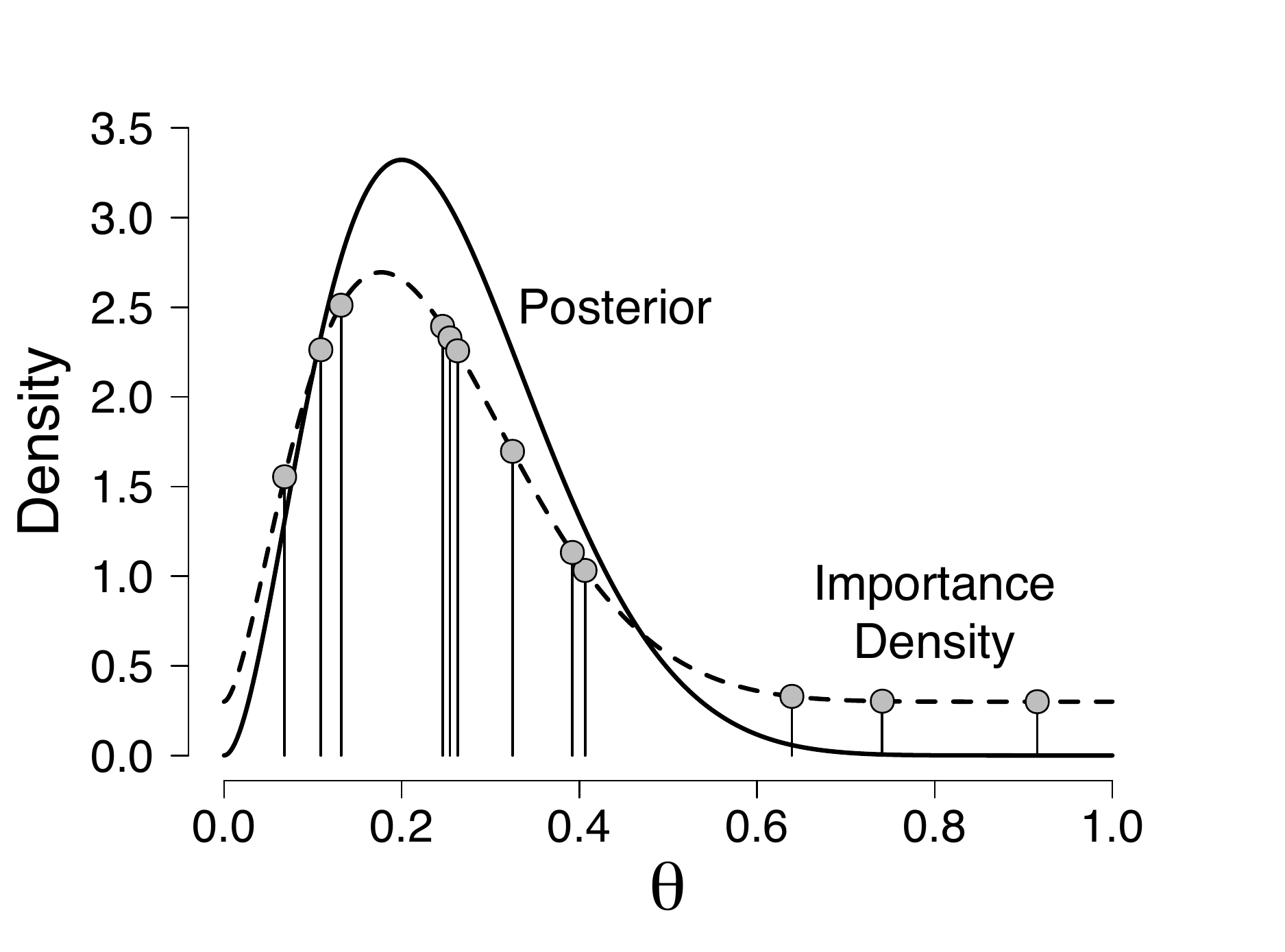}
    \caption{Illustration of the importance sampling estimator for the beta-binomial model. The dashed line represents our beta mixture importance density and the solid gray line represents the posterior distribution that was obtained after having observed 2 correct responses out of 10 trials. The gray dots represent the 12 samples $\{ \tilde \theta_1, \tilde \theta_2, \ldots, \tilde \theta_{12} \}$ randomly drawn from our beta mixture importance density. Available at \url{https://tinyurl.com/yc7ho7hr} under CC license \url{https://creativecommons.org/licenses/by/2.0/}.}
	\label{importanceSample}
\end{figure}

\noindent
Using a mixture weight on the uniform component of $\gamma = 0.30$---a choice that was made to ensure that, visually, the tails of the importance density are clearly thicker than the tails of the posterior distribution---we obtain the following importance density: $\gamma \times \text{Beta}(\theta; \; 1, 1) + (1-\gamma) \times \text{Beta}(\theta; \; \hat \alpha, \hat \beta) =  .3 + .7 \; \text{Beta}(\theta; \; 2.721, 9.006)$. This importance density is represented by the dashed line in Figure \ref{importanceSample}. The figure also shows the posterior distribution (solid line). As is evident from the figure, the beta mixture importance density resembles the posterior distribution, but has fatter tails.

In general, it is advised to choose the mixture weight on the uniform component $\gamma$ small enough to make the estimator efficient, yet large enough to produce fat tails to stabilize the estimator. A suitable mixture weight can be realized by gradually minimizing the mixture weight and investigating whether stability is still guaranteed (i.e., robustness analysis). 

Drawing $N = 12$ samples for $\theta$ from our beta mixture importance density results in:

\begin{align*}
\{ \tilde \theta_1, \tilde \theta_2, \ldots, \tilde \theta_{12} \} =&  \{0.11, 0.07, 0.32, 0.25, 0.41, 0.39, 0.25, 0.13, 0.64, 0.26, 0.74, 0.92\}.
\end{align*}

\noindent
These samples are represented by the gray dots in Figure \ref{importanceSample}.

The final step is to compute the average adjusted likelihood for the 12 samples using Equation~\ref{Eq:ISE}. This yields the importance sampling estimate of the marginal likelihood as:

\begin{align*} 
 \hat p_2(k = 2 \mid n = 10) &= \cfrac{1}{12} \sum_{i = 1}^{12}
         \cfrac{p(k = 2\mid n = 10,  \tilde \theta_i) \; p(\tilde \theta_i)}
              {.3 + .7 \; \text{Beta}(\tilde \theta_i; \;  2.721, 9.006) } \\
              \\
         &= \cfrac{1}{12} 
              \left(\cfrac{\left( {10 \atop 2} \right ) 0.11^2 (1 - 0.11)^8 \times 1}
         {.3 + .7 \; \text{Beta}(0.11; \; 2.721, 9.006)}   
        + \ldots +                              
        \cfrac{\left( {10 \atop 2} \right ) 0.92^2 (1 - 0.92)^8 \times 1}
         {.3 + .7 \; \text{Beta}(0.92; \; 2.721, 9.006) }\right)\\
         & =  \cfrac{1}{12} \binom{10}{2} (0.0021 + \ldots + 7.3 \times 10^{-9}) \\
         & =  0.0827.
\end{align*}

\subsection{Method 3: The Generalized Harmonic Mean Estimator of the Marginal Likelihood}

Just as the importance sampling estimator, the generalized harmonic mean estimator focuses on regions of the parameter space where the integrand of Equation~\ref{Eq:ML} is large by using an importance density $g_{IS}(\theta)$ \cite{GelfandKey1994}.\footnote{Note that the generalized harmonic mean estimator is a more stable version of the harmonic mean estimator \cite{NewtonRaftery1996}. A problem of the harmonic mean estimator is that it is dominated by the samples that have small likelihood values.} However, in contrast to the importance sampling estimator, the generalized harmonic mean estimator requires an importance density with thinner tails for an analogous reason as in importance sampling.

To derive the generalized harmonic mean estimator, also known as reciprocal importance sampling estimator \cite{fruhwirth2004estimating}, we use the following identity: 

\begin{equation*}
  \begin{split}
    \cfrac{1}{p(y)} 
    &= \int \cfrac{1}{p(y)} \; g_{IS}(\theta) \; \mathrm{d}\theta 
    = \int \cfrac{p(\theta \mid y)}{p(y\mid \theta) p(\theta)} \; g_{IS}(\theta) \; \mathrm{d}\theta 
    = \int \cfrac{g_{IS}(\theta) }{p(y\mid \theta) p(\theta)} \; p(\theta \mid y) \; \mathrm{d}\theta \\
    &= \mathbb{E}_\text{post}
                           \left(\cfrac{g_{IS}(\theta)}{p(y\mid \theta) \;
                           p(\theta)}\right)  .
  \end{split}
\end{equation*}

\noindent
Rewriting results in:

\begin{align*}
               p(y) &= \left( \mathbb{E}_\text{post}
                           \left(\cfrac{g_{IS}(\theta)}{p(y\mid \theta)
                           p(\theta)}\right) \right) ^{-1},
\end{align*}

\noindent
which is used to define the generalized harmonic mean estimator $\hat p_{3}(y)$ \cite{GelfandKey1994} as follows:

\begin{align} \label{Eq:GHME}
        \hat p_{3}(y) = \left( \cfrac{1}{N} \sum_{j = 1}^N 
     \cfrac{\overbrace{g_{IS}(\theta^*_j)}^\text{importance density}}{\underbrace{p(y\mid \theta^*_j)}_\text{likelihood}
                 \; \underbrace{p(\theta^*_j)}_\text{prior}}
        \right) ^{-1}, \; \; \underbrace{\theta^*_j \sim p(\theta\mid y) \; .}_{\substack{\text{samples from the}\\ \text{posterior distribution}}}
 \end{align}

Note that the generalized harmonic mean estimator---in contrast to the importance sampling estimator---evaluates samples from the posterior distribution. In addition, note that the ratio in Equation~\ref{Eq:GHME} is the reciprocal of the ratio in Equation~\ref{Eq:ISE}; this explains why the importance density for the generalized harmonic mean estimator should have thinner tails than the posterior distribution in order to avoid inflation of the ratios that are part of the summation displayed in Equation~\ref{Eq:GHME}.
Thus, in the case of the generalized harmonic mean estimator, a suitable importance density  should (1) have thinner tails than the posterior distribution \cite{NewtonRaftery1996, Dicicciokass1997}, and as in importance sampling, it should (2) be easy to evaluate; (3) have the same domain as the posterior distribution; and (4) closely resemble the posterior distribution.

\subsubsection{Running example}
To obtain the generalized harmonic mean estimate of the marginal likelihood in our running example, we need to choose a suitable importance density. In our running example, an importance density that fulfills the four above mentioned desiderata can be obtained by following four steps: First, we draw $N = 12$ samples from the posterior distribution. Reusing the samples from the last section, we obtain:

\begin{align*}
\{ \theta^*_1, \theta^*_2, \ldots, \theta^*_{12} \} =& \{0.22, 0.16, 0.09, 0.35, 0.06, 0.27, 0.26, 0.41, 0.20, 0.43, 0.21, 0.12\}.
\end{align*}

Second, we probit-transform all posterior samples (i.e., $\xi^*_j = \Phi^{-1}(\theta^*_j), \; \text{with} \; j \in \{1, 2, \ldots, 12\}$).\footnote{Other transformation are conceivable (e.g., logit transformation).} The result of this transformation is that the samples range across the entire real line instead of the $(0, 1)$ interval only. We obtain:

\begin{align*}
\{ \xi^*_1, \xi^*_2, \ldots, \xi^*_{12}\} =& \{-0.77, -0.99, -1.34, -0.39, -1.55, -0.61, -0.64, -0.23, -0.84, -0.18, \\ & \; -0.81,
-1.17\}.
\end{align*}

\begin{figure}[!bt]
	\centering
    \includegraphics[width=0.8\textwidth]{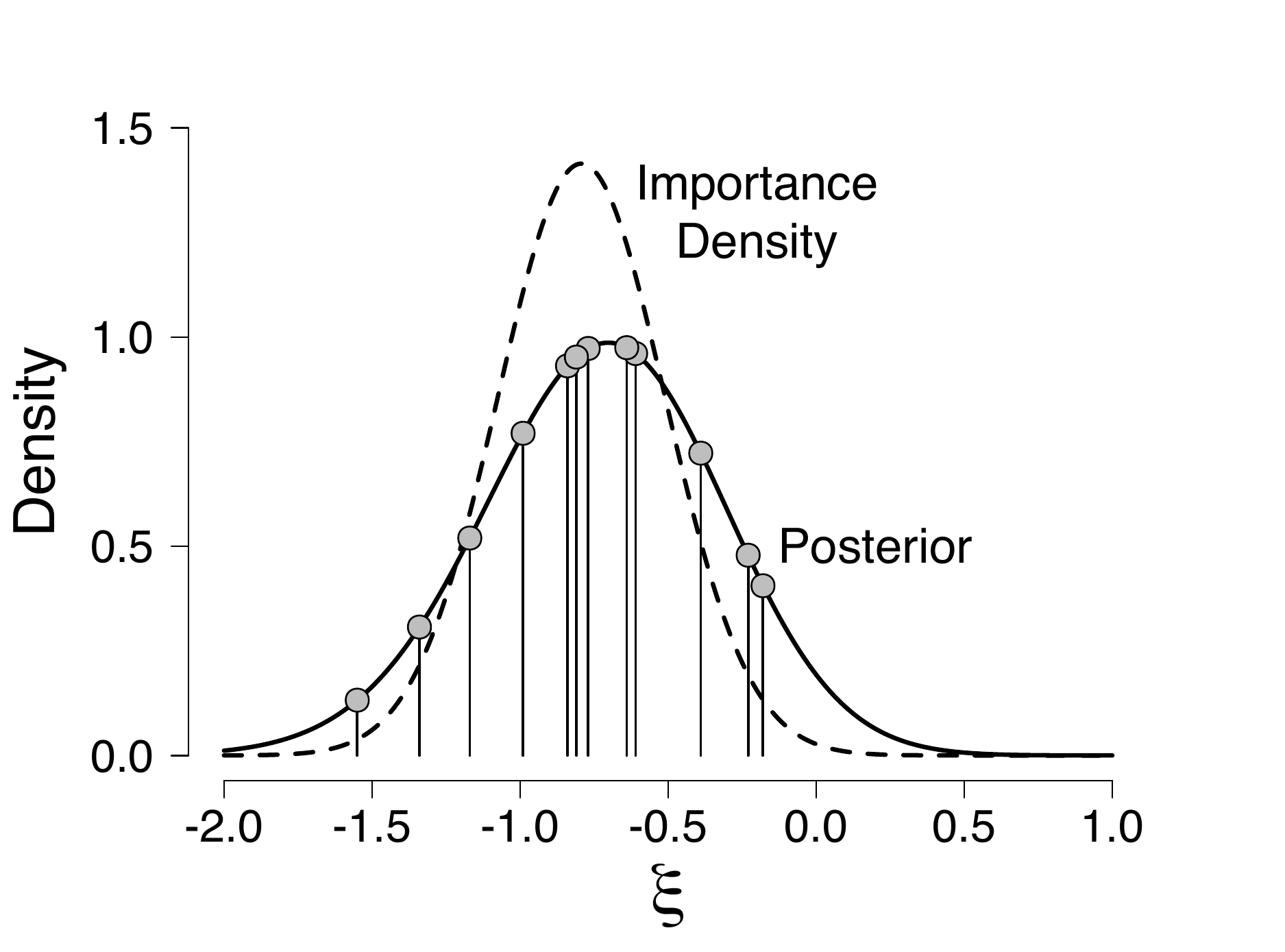}%{images/GHMSample.eps}
	\caption{Illustration of the generalized harmonic mean estimator for the beta-binomial model. The solid line represents the probit-transformed $\text{Beta}(3, 9)$ posterior distribution that was obtained after having observed 2 correct responses out of 10 trials, and the dashed line represents the importance density $\mathcal{N}\left(\xi ; \; \mu = -0.793, \sigma = 0.423/1.5\right)$. The gray dots represent the 12 probit-transformed samples $\{\xi^*_1, \xi^*_2, \ldots, \xi^*_{12} \}$ randomly drawn from the $\text{Beta}(3, 9)$ posterior distribution. Available at \url{https://tinyurl.com/yazgk8kj} under CC license \url{https://creativecommons.org/licenses/by/2.0/}.}
	\label{GHMSample}
\end{figure}

\noindent
These probit-transformed samples are represented by the gray dots in Figure \ref{GHMSample}.

Third, we search for the normal distribution that provides the best fit to the probit-transformed posterior samples $\xi^*_j$. Using the method of moments, we obtain as estimates $\hat \mu = -0.793$ and $\hat \sigma = 0.423$. Note that the choice of a normal importance density justifies step 2; the probit transformation (or an equivalent transformation) was required to match the range of the posterior distribution to the one of the normal distribution.

Finally, as importance density we choose a normal distribution with mean $\hat \mu = -0.793$ and standard deviation $\hat \sigma = 0.423 / 1.5$. This additional division by 1.5 is to ensure thinner tails of the importance density than of the probit-transformed posterior distribution (for a discussion of alternative importance densities see \citeNP{Dicicciokass1997}). We decided to divide $\hat \sigma$ by $1.5$ for illustrative purposes only. Our importance density is displayed in Figure \ref{GHMSample} (dashed line) together with the probit-transformed posterior distribution (solid line).

The generalized harmonic mean estimate can now be obtained using either the original posterior samples $\theta^*_j$ or the probit-transformed samples $\xi^*_j$. Here we use the latter ones (see also \citeNP{overstall2010default}). Incorporating our specific importance density and a correction for having used the probit-transformation, Equation~\ref{Eq:GHME} becomes:\footnote{A detailed explanation is provided in the appendix. Note that using the original posterior samples $\theta^*_j$ would involve transforming the importance density (e.g., the normal density on $\xi$) to the $(0, 1)$ interval.}
 
\begin{align} \label{Eq:GHMEprobit}
\begin{split}
        \hat p_{3}(y) &= \left( \cfrac{1}{N} \mathlarger{\sum}_{j = 1}^N 
     \cfrac{\overbrace{\frac{1}{\hat \sigma} \phi \left ( \frac{\xi^*_j - \hat \mu}{\hat \sigma}\right ) }^\text{importance density}}{\underbrace{p \left ( y\mid \Phi \left ( \xi^*_j \right ) \right )}_\text{likelihood}
                 \; \underbrace{ \phi\left ( \xi^*_j \right ) }_\text{prior}}
        \right) ^{-1}, \; \; \underbrace{\xi^*_j = \Phi^{-1}(\theta^*_j) \; \text{and} \; \theta^*_j \sim p(\theta\mid y) \; .}_{\substack{\text{probit-transformed samples}\\ \text{from the posterior distribution}}}     
\end{split}        
 \end{align}

\par
For our beta-binomial model, we now obtain the generalized harmonic mean estimate of the marginal likelihood as:

\begin{align*} 
 \hat p_{3}(k = 2 \mid n = 10) &=  \left( \cfrac{1}{12} \sum_{j = 1}^{12}
        \cfrac{\frac{1}{0.423/1.5} \, \phi \left ( \frac{\xi_j^* + 0.793}{0.423/1.5}\right )}{p(k = 2 \mid n = 10, \Phi(\xi^*_j)) \;
        \phi(\xi^*_j)}\right)^{-1} \\  
      &= \left( \cfrac{1}{12} 
         \left(\cfrac{\frac{1}{0.423/1.5} \, \phi \left ( \frac{-0.77 + 0.793}{0.423/1.5}\right )} % g(theta)
         {\binom{10}{2} 0.22^2 (1 - 0.22)^8 \; \phi(-0.77)}    % likelihood * prior
          + \ldots + 
          \cfrac{\frac{1}{0.423/1.5} \, \phi \left ( \frac{-1.17 + 0.793}{0.423/1.5}\right )}      % g(theta)
         {\binom{10}{2} 0.12^2 (1 - 0.12)^8 \; \phi(-1.17)    % likelihood * prior
         }\right) \right) ^{-1} \\
      &=  \left( \cfrac{1}{12} \; \cfrac{1}{\binom{10}{2}} \; \left( 716.89 + \ldots +
         555.50 \right) \right)^{-1} \\
      &=   0.092.
\end{align*}

\subsection{Method 4: The Bridge Sampling Estimator of the Marginal Likelihood}

As became evident in the last two sections, both the importance sampling estimator and the generalized harmonic mean estimator  impose strong constraints on the tail behavior of the importance density relative to the posterior distribution to guarantee a stable estimator. Such requirements can make it difficult to find a suitable importance density, especially when a high-dimensional posterior is considered. The bridge sampler, on the other hand,  alleviates such requirements (e.g., \citeNP{fruhwirth2004estimating}).

Originally, bridge sampling was developed to directly estimate the Bayes factor, that is, the ratio of the marginal likelihoods of two models $ \mathcal{M}_1$ and $ \mathcal{M}_2$ (e.g., \citeNP{Jeffreys1961BF, kass1995bayes}). However, in this tutorial, we use a version of bridge sampling that allows us to approximate the marginal likelihood of a \emph{single} model (for an earlier application see for example \citeNP{overstall2010default}).  This version is based on the following identity:

\begin{align}
\label{E:BS1model}
        1&= \cfrac{\int p(y\mid \theta ) \;p(\theta) 
                        \; h(\theta )
                        \; g(\theta ) \;
                \mathrm{d}\theta}
                    {\int p(y\mid \theta)
                        \; p(\theta) 
                        \; h(\theta )
                        \; g(\theta ) \;
                \mathrm{d}\theta} \; ,
\end{align}

\noindent
where  $g(\theta)$ is the so-called proposal distribution and $h(\theta)$ the so-called bridge function. Multiplying both sides of Equation~\ref{E:BS1model} by the marginal likelihood 
$p(y)$ results in:

\begin{align*}
         p(y)&= 
                  \cfrac{\int p(y\mid \theta ) 
                        \; p(\theta )
                        \; h(\theta )
                        \; g(\theta ) \;
                        \mathrm{d}\theta}
                        {\displaystyle\int \cfrac{p(y\mid \theta )
                                \; p(\theta)}
                                {p(y)}
                        \; h(\theta )
                        \; g(\theta ) \;
                \mathrm{d}\theta} \; =  
                \cfrac{\int
                        {p(y\mid \theta)}\; {p(\theta)}
                        \; 
                         h(\theta)
                        \overbrace{g(\theta)}^{\substack{\text{proposal}\\\text{distribution}}}
                        \mathrm{d}\theta}
                        {\int
                        {h(\theta)} \,
                        {g(\theta)}
                        \;\; \; \; \;\; \; \; \; \;\; \; \; 
                        \underbrace{p(\theta \mid y)}_{\substack{\text{posterior}\\\text{distribution}}}
                        \mathrm{d}\theta 
                       }  \\
                       \\
&=   \cfrac{\mathbb{E}_{g(\theta)} \left[
                          p(y\mid \theta )  
                        \;p(\theta ) 
                        \; h(\theta )\right] }
                      {\mathbb{E}_\text{post}\left[h(\theta ) \; g(\theta )\right]} \ .                         
\end{align*} 

The marginal likelihood can now be approximated using:

\begin{align} 
  \label{eq:f7}
       \hat p(y)
             &= \cfrac{\frac{1}{N_2} \sum_{i = 1}^{N_2} 
                             p(y\mid \tilde \theta_i) 
                             \; p(\tilde \theta_i)
                             \; h(\tilde \theta_i)}
               {\frac{1}{N_1} \sum_{j = 1}^{N_1} 
                             h(\theta^*_j)
                             \; g(\theta^*_j)}, \ \ \underbrace{\tilde \theta_i \sim g(\theta)}_{\substack{\text{samples from the}\\\text{proposal distribution}}}, \ \ \underbrace{\theta^*_j \sim p(\theta \mid y) \; .}_{\substack{\text{samples from the}\\\text{posterior distribution}}}
\end{align}

Equation~\ref{eq:f7} illustrates that we need samples from both the proposal distribution and the posterior distribution to obtain the bridge sampling estimate for the marginal likelihood. However, before we can apply Equation~\ref{eq:f7} to our running example, we have to discuss how we can obtain a suitable proposal distribution and bridge function. Conceptually, the proposal distribution is similar to an importance density, should resemble the posterior distribution, and should have sufficient overlap with the posterior distribution. According to \citeA{overstall2010default}, a convenient proposal distribution is often a normal distribution with its first two moments chosen to match those of the posterior distribution.  In our experience, this choice for the proposal distribution works well for a wide range of scenarios. However, this proposal distribution might produce unstable estimates in case of high-dimensional posterior distributions that clearly do not follow a multivariate normal distribution. In such a situation, it might be advisable to consider more sophisticated versions of bridge sampling (e.g., \citeNP{fruhwirth2004estimating, Meng2002, wang2016warp}). 

\subsubsection{Choosing the optimal bridge function}
In this tutorial we use the bridge function defined as \cite{MengWong1996}:

\begin{align}
  \label{Eq:BridgeFnc}
  h(\theta) = C \cdot \frac{1}{s_1 p(y\mid \theta)p(\theta) +
s_2 p(y)g(\theta)} \; ,
\end{align}

\noindent
where $s_1 = \frac{N_1}{N_2 + N_1}$, $s_2 = \frac{N_2}{N_2 + N_1}$, and $C$ a constant; its particular value is not required because $h(\theta)$ is part of both the numerator and the denominator of Equation~\ref{eq:f7}, and therefore the constant $C$ cancels. This particular bridge function is referred to as the ``optimal bridge function'' because \citeA[p.~837]{MengWong1996} proved that it minimizes the relative mean-squared error  (Equation~\ref{E:RE2}).

Equation~\ref{Eq:BridgeFnc} shows that the optimal bridge function depends on the marginal likelihood $p(y)$ which is the very entity we want to approximate. We can resolve this issue by applying an iterative scheme that updates an initial guess of the marginal likelihood until the estimate of the marginal likelihood has converged according to a predefined tolerance level. To do so, we insert the expression for the optimal bridge function (Equation~\ref{Eq:BridgeFnc}) in Equation~\ref{eq:f7} \cite{MengWong1996}. The formula to approximate the marginal likelihood on iteration $t + 1$ is then specified as follows:

\begin{align} 
  \label{E:BSIter}
  \begin{split}
        &\hat p(y)^{(t + 1)} = 
        \cfrac{\cfrac{1}{N_2} \mathlarger{\sum}_{i = 1}^{N_2}
                \cfrac{p(y\mid \tilde \theta_i)p(\tilde \theta_i)}
             {s_1 p(y\mid \tilde \theta_i) p(\tilde \theta_i) + 
              s_2 \hat p(y)^{(t)} g(\tilde \theta_i)}}
              {\cfrac{1}{N_1} \mathlarger{\sum}_{j = 1}^{N_1} \cfrac{
                        g(\theta^*_j)}
                        {s_1 p(y\mid \theta^*_j) p(\theta^*_j) + 
                        s_2 \hat p(y)^{(t)} g(\theta^*_j)}} \; , \\ 
                        \\
                        &  \hspace{2.6cm} \underbrace{\tilde \theta_i \sim g(\theta)}_{\substack{\text{samples from the}\\\text{proposal distribution}}}, \ \ \underbrace{\theta^*_j \sim p(\theta\mid y)}_{\substack{\text{samples from the}\\\text{posterior distribution}}},
  \end{split}
\end{align}

\begin{sloppypar}
\noindent
where $\hat p(y)^{(t)}$ denotes the estimate of the marginal likelihood on
iteration $t$ of the iterative scheme.  Note that Equation~\ref{E:BSIter} illustrates why bridge sampling is robust to the tail behavior of the proposal distribution relative to the posterior distribution;  the difference to the importance sampling and generalized harmonic mean estimator is that, in the case of the bridge sampling estimator, samples from the tail region cannot inflate individual summation terms and thus dominate the estimate.
To illustrate this, we consider what happens to the bridge sampling estimator, the importance sampling estimator, and the generalized harmonic mean estimator in case (1) the proposal/importance distribution has fatter tails than the posterior distribution, and (2) the proposal/importance distribution has thinner tails than the posterior distribution \cite<see also>{fruhwirth2004estimating}. 
Specifically, we look at a single term in the respective sums and consider the limit of that term as we move further and further out in the tails. This is insightful since a single term can have a lasting effect on the estimator (e.g., in case a single term in a sum is very large or even infinite).

In case (1) (i.e., the proposal/importance distribution has fatter tails than the posterior), the ratio in the importance sampling estimator (i.e., Equation~\ref{Eq:ISE}) goes to zero as we move further out in the tails. Since samples in the tails may only be obtained occasionally and a zero term in the sum does not inflate the estimate this is not a reason for concern. In contrast, when we consider the ratio in the generalized harmonic mean estimator (i.e., Equation~\ref{Eq:GHME}), we see that the ratio goes to infinity as we move further out in the tails. Even if this occurs only very rarely, this is an issue since the resulting value will dominate the estimate. Consequently, the resulting estimator may have a large variance since samples from the tail regions may be obtained only occasionally across repeated applications. For the bridge sampling estimator (i.e., Equation~\ref{E:BSIter}), we need to consider the ratio in the numerator and denominator. The ratio in the numerator will go to zero and the ratio in the denominator will go to $\frac{1}{s_2 \, \hat p(y)^{(t)}}$. Hence, both of these ratios are bounded and will not inflate the two sums, hence also not the resulting estimate.

In case (2) (i.e., the proposal/importance distribution has thinner tails than the posterior), the ratio in the importance sampling estimator (i.e., Equation~\ref{Eq:ISE}) goes to infinity as we move further out in the tails, inflating the estimate. In contrast, when we consider the ratio in the generalized harmonic mean estimator (i.e., Equation~\ref{Eq:GHME}), we see that the ratio goes to zero. As explained above, this is not a reason for concern. These considerations explain why in importance sampling, the importance distribution should have fatter tails than the posterior whereas for the generalized harmonic mean estimator, it should have thinner tails. For the bridge sampling estimator (i.e., Equation~\ref{E:BSIter}), the ratio in the numerator will go to $1/s_1$ and the ratio in the denominator will go to zero. Again, both of these ratios are bounded making the bridge sampling estimator more robust to the tail behavior than the other two estimators. This of course assumes that not all terms in the denominator (for case (2)) and the numerator (for case (1)) will be zero, that is, the proposal and the posterior distribution have sufficient overlap.
 In the extreme scenario of no overlap the bridge sampling estimate is not defined because both sums  of Equation~\ref{E:BSIter} would be zero.

Extending the numerator of the right side of Equation~\ref{E:BSIter} with $\cfrac{1/g(\tilde \theta_i)}{1/g(\tilde \theta_i)}$, and the denominator with $\cfrac{1/g(\theta^*_j)}{1/g(\theta^*_j)}$, and subsequently defining $l_{1,j} := \cfrac{p(y\mid \theta^*_j) p(\theta^*_j)}{g(\theta^*_j)}$ and $l_{2,i} := \cfrac{p(y\mid \tilde \theta_i) p(\tilde \theta_i)}{g(\tilde \theta_i)}$, we obtain the formula for the iterative scheme of the bridge sampling estimator $\hat p_4(y)^{(t + 1)}$ at iteration $t+1$ \cite[p.~837]{MengWong1996}. 
\end{sloppypar}

\begin{align} 
  \label{eq:f9}
  \begin{split}
       \hat p_4(y)^{(t + 1)} &= 
       \cfrac{\cfrac{1}{N_2} \mathlarger{\sum}_{i = 1}^{N_2} 
               \cfrac{p(y\mid \tilde \theta_i)p(\tilde \theta_i)}
                       {s_1 p(y\mid \tilde \theta_i) p(\tilde \theta_i) + 
                       s_2 \hat p_4(y)^{(t)} g(\tilde \theta_i)}
                     \; \cfrac{1/g(\tilde \theta_i)}{1/g(\tilde \theta_i)}}
                       {\cfrac{1}{N_1} \mathlarger{\sum}_{j = 1}^{N_1} \cfrac{
                       g(\theta^*_j)}
                       {s_1 p(y\mid \theta^*_j) p(\theta^*_j) + 
                       s_2 \hat p_4(y)^{(t)} g(\theta^*_j)} \;
            \cfrac{1/g(\theta^*_j)}{1/g(\theta^*_j)}} \\
            \\
         &= \cfrac{\cfrac{1}{N_2} \mathlarger{\sum}_{i = 1}^{N_2} 
                      \cfrac{l_{2,i}}{s_1 l_{2,i} + s_2 \hat p_4(y)^{(t)}}}
                     {\cfrac{1}{N_1} \mathlarger{\sum}_{j = 1}^{N_1}
                      \cfrac{1}{s_1 l_{1,j} + s_2 \hat p_4(y)^{(t)}}} \; , \ \ \underbrace{\tilde \theta_i \sim g(\theta)}_{\substack{\text{samples from the}\\\text{proposal distribution}}}, \ \ \underbrace{\theta^*_j \sim p(\theta\mid y)}_{\substack{\text{samples from the}\\\text{posterior distribution}}} .\\
  \end{split}
\end{align}

Equation~\ref{eq:f9} suggests that, in order to obtain the bridge sampling estimate of the marginal likelihood, a number of requirements need to be fulfilled. First, we need $N_2$ samples from the proposal distribution $g(\theta)$ and $N_1$ samples from the posterior distribution $p(\theta | y)$. Second, for all $N_2$ samples from the proposal distribution, we have to evaluate $l_{2,i}$. This involves obtaining the value of the unnormalized posterior (i.e., the product of the likelihood times the prior) and of the proposal distribution for all samples. Third, we evaluate $l_{1,j}$ for all $N_1$ samples from the posterior distribution. This is analogous to evaluating $l_{2,i}$. Fourth, we have to determine the constants $s_1$ and $s_2$ that only depend on $N_1$ and $N_2$. Fifth, we need an initial guess of the marginal likelihood $\hat p_4(y)$. Since some of these five requirements can be obtained easier than others, we will point out possible challenges.

A first challenge is that using a suitable proposal distribution may involve transforming the posterior samples. Consequently, we have to determine how the transformation affects the definition of the bridge sampling estimator for the marginal likelihood (Equation~\ref{eq:f9}). 

A second challenge is how to use the $N_1$ samples from the posterior distribution. One option is to use all $N_1$ samples for both fitting the proposal distribution and for computing the bridge sampling estimate. However, \citeA{overstall2010default} showed that such a procedure may result in an underestimation of the marginal likelihood. To obtain more reliable estimates they propose to divide the posterior samples in two parts; the first part is used to obtain the best-fitting proposal distribution, and the second part is used to compute the bridge sampling estimate. Throughout this tutorial, we use two equally large parts. In the remainder we therefore state that we draw $2 N_1$ samples from the posterior distribution. The first $N_1$ of the total of $2 N_1$ samples are used for fitting the proposal distribution and the remaining $N_1$ samples are used in the iterative scheme (i.e., Equation~\ref{eq:f9}).\footnote{In case the posterior samples are obtained via MCMC sampling using multiple chains, we use the first half of the iterations per chain for fitting the proposal distribution and the second half of the iterations per chain for the iterative scheme.}

To summarize, the discussion of the requirements and challenges encountered in bridge sampling illustrated that the bridge sampling estimator imposes less strict requirements on the proposal distribution than the importance sampling and generalized harmonic mean estimator and allows for an almost automatic application due to the default choice of the bridge function.\footnote{For an explanation of where the name ``bridge'' comes from see \url{https://osf.io/9jzm3/}.}
 
\subsubsection{Running example}

\begin{figure}[!bt]
	\centering
    \includegraphics[width=0.8\textwidth]{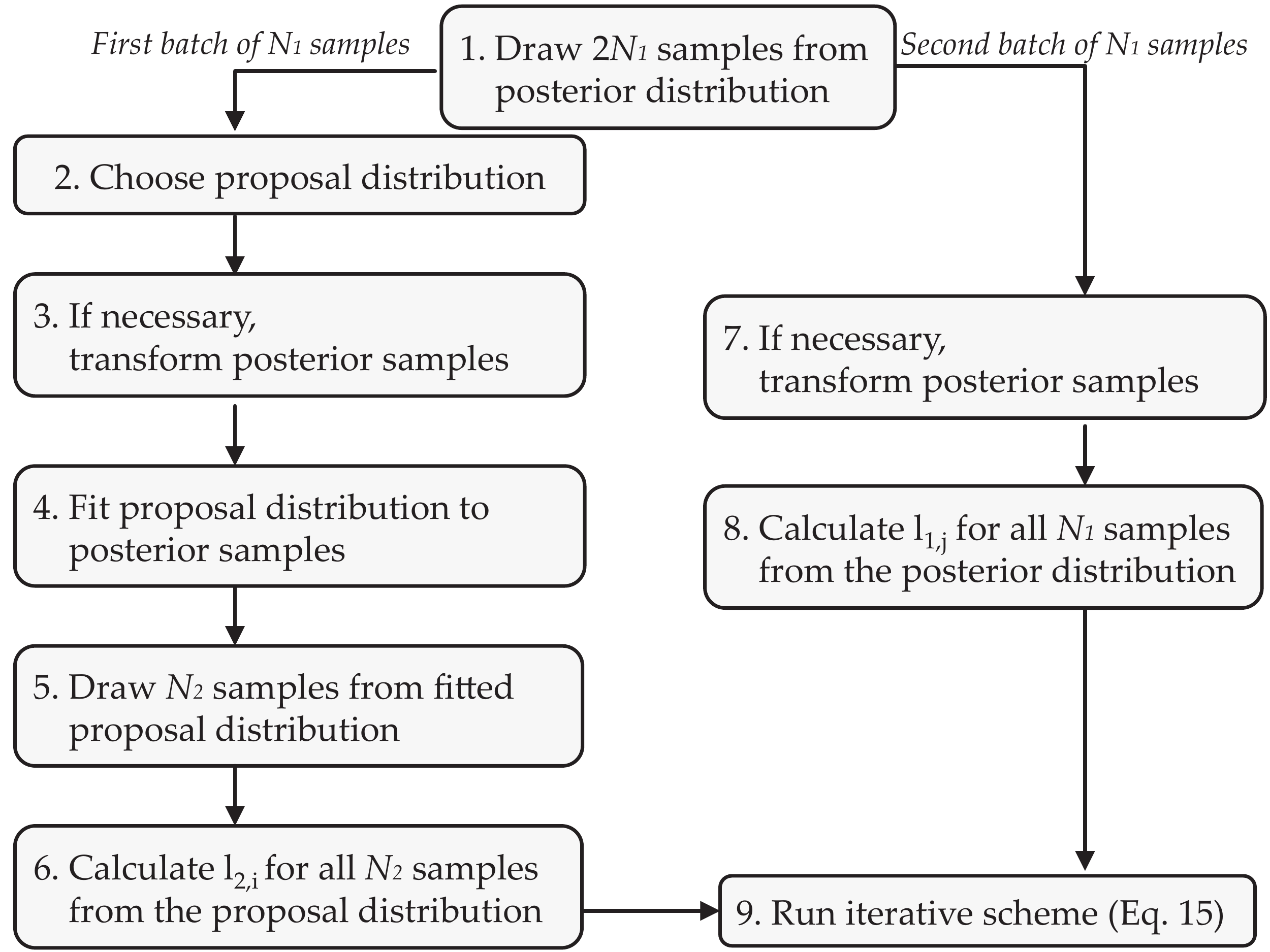}
    \caption{Schematic illustration of the steps involved in obtaining the bridge sampling estimate of the marginal likelihood. Available at \url{https://tinyurl.com/y7b2kze7} under CC license \url{https://creativecommons.org/licenses/by/2.0/}.} 
	\label{F:algorithm}
\end{figure}
 
To obtain the bridge sampling estimate of the marginal likelihood in the beta-binomial example, we follow the eight steps illustrated in Figure~\ref{F:algorithm}:
\begin{enumerate}
\item \emph{We draw $2 N_1 = 24$ samples from the $\text{Beta}(3,9)$ posterior distribution for $\theta$.} \\
We obtain the following sample of 24 values:
\begin{align*}
\{ \theta^*_1, \theta^*_2, \ldots, \theta^*_{24} \}  =& \{0.22, 0.16, 0.09, 0.35, 0.06, 0.27, 0.26, 0.41, 0.20, 0.43, 0.21, 0.12, \\ 
& \; 0.15, 0.21, 0.24, 0.18, 0.12, 0.22, 0.15, 0.22, 0.23, 0.26, 0.29, 0.28\}.
\end{align*}
Note that the first $12$ samples equal the ones used in the last section, whereas the last 12 samples were obtained from drawing again 12 values from the $\text{Beta}(3,9)$ posterior distribution for $\theta$.
\item \emph{We choose a proposal distribution.} \\
Here we opt for an approach that can be easily generalized to models with multiple parameters and select a normal distribution as the proposal distribution $g(\theta)$.\footnote{There exist several candidates for the proposal distribution. Alternative proposal distributions are, for example, the importance density that we used for the importance sampling estimator or for the generalized harmonic mean estimator, or the analytically derived $\text{Beta}(3,9)$ posterior distribution.} 
\item \emph{We transform the first batch of $N_1$ posterior samples.} \\
Since we use a normal proposal distribution, we have to transform the posterior samples from the rate scale to the real line so that the range of the posterior distribution matches the range of the proposal distribution. This can be achieved by probit-transforming the posterior samples, that is, $\xi_j^*=\Phi^{-1}(\theta^*_j)$ with $j \in \{1, 2, \ldots, 12\}$. We obtain: 
\begin{align*}
\{ \xi^*_1, \xi^*_2, \ldots, \xi^*_{12}\} =&  \{ -0.77, -0.99, -1.34, -0.39, -1.55, -0.61, -0.64, -0.23, -0.84, -0.18, \\ & \;-0.81, -1.17\}.
\end{align*}
\item \emph{We fit the proposal distribution to the first batch of $N_1$ probit-transformed posterior samples.}\\
We use the method of moment estimates $\hat \mu = -0.793$ and $\hat \sigma = 0.423$ from the first batch of $N_1$ probit-transformed posterior samples to obtain our proposal distribution $g(\xi ; \mu = -0.793, \sigma = 0.423) = \frac{1}{0.423} \, \phi \left ( \frac{\xi + 0.793}{0.423}\right )$. 
\item \emph{We draw $N_2$ samples from the proposal distribution.} \\
We obtain:
\begin{align*}
\{ \tilde \xi_1, \tilde \xi_2, \ldots, \tilde \xi_{12} \} =&
\{ -1.11, -0.63, -1.48, -0.59, -0.48, -0.69, -0.74, -0.51, -0.82,
\\ & \; -1.54, -0.76, -0.96\}.
\end{align*}

\item \emph{We calculate $l_{2,i}$ for all $N_2$ samples from the proposal distribution.}\\
This step involves assessing the value of the unnormalized posterior and the proposal distribution for all $N_2$ samples from the proposal distribution. As in the running example for the generalized harmonic mean estimator, we obtain the unnormalized posterior as: $p \left ( k=2 \mid n=10, \Phi \left (\tilde \xi_i \right ) \right ) \phi\left (\tilde \xi_i \right )$, where $\phi\left (\tilde \xi_i \right )$ comes from using the change-of-variable method (see running example for the generalized harmonic mean estimator and the appendix for details). Thus, as in the case of the generalized harmonic mean estimator, the uniform prior on $\theta$ translates to a standard normal prior on $\xi$. The values of the proposal distribution can easily be obtained (for example using the R software). 
\item \emph{We transform the second batch of $N_1$ posterior samples.} \\ 
As in step 2, we use the probit transformation and obtain:
\begin{align*}
\{ \xi^*_{13}, \xi^*_{14}, \ldots, \xi^*_{24}\} =& \{-1.04, -0.81, -0.71, -0.92, -1.17, -0.77, -1.04, -0.77, -0.74, \\ & \;  -0.64, -0.55, -0.58\}.
\end{align*}
\item{ \emph{We calculate $l_{1,j}$ for the second batch of $N_1$ probit-transformed samples from the posterior distribution.}}\\
This is analogous to step 6.
\item \emph{We run the iterative scheme (Equation~\ref{eq:f9}) until our predefined tolerance criterion is reached.} \\
As tolerance criterion we choose ${|\hat p_4(k = 2  \mid n = 10)^{(t + 1)} - \hat p_4(k = 2  \mid n = 10)^{(t)} |} \, / $ ${\hat p_4(k = 2  \mid n = 10)^{(t + 1)}} \leq 10^{-10}$. This requires an initial guess for the marginal likelihood $\hat p_4(k = 2  \mid n = 10)^{(0)}$ which we set to 0.\footnote{A better initial guess can be obtained from, for example, the importance sampling estimator or the generalized harmonic mean estimator explained in the previous sections. In our experience, however, usually the exact choice of the initial value does not seem to influence the convergence of the bridge sampler much.}
\end{enumerate} 

The simplicity of the beta-binomial model allows us to calculate the bridge sampling estimate by hand. To determine $\hat p_4(y)^{(t+1)}$ according to Equation~\ref{eq:f9}, we need to calculate the constants $s_1$ and $s_2$. Since $N_1 = N_2 = 12$, we obtain: $s_1 = s_2 = {N_2} / ({N_2 + N_1}) = 0.5$. In addition, we need to calculate $l_{2,i}$ ($i \in \{1, 2, \ldots, 12\}$) for all samples from the proposal distribution, and $l_{1,j}$ ($j \in \{1, 2, \ldots, 12\}$) for the second batch of the probit-transformed samples from the posterior distribution. Here we show how to calculate $l_{2,1}$ and $l_{1,1}$ using the first sample from the proposal distribution and the first sample of the second batch of the posterior samples, respectively:

\begin{align*}
        l_{2,1}
         &=  \cfrac{p(k\mid n, \Phi(\tilde \xi_1)) \phi(\tilde \xi_1)}{g(\tilde \xi_1)}
         = \left(\cfrac{\left( {10 \atop 2} \right ) 0.13^2 (1 - 0.13)^8  \cdot 0.22}
                 {\frac{1}{0.423} \, \phi \left ( \frac{-1.11 +  0.793}{0.423}\right )}\right)   
          =  0.077, \\
\end{align*}

\begin{align*}
        l_{1,1}
        &= \cfrac{p(k\mid n,  \Phi(\xi^*_{13})) 
        \phi(\xi^*_{13})}{g(\xi^*_{13})} 
         = \left(\cfrac{\left( {10 \atop 2} \right ) 0.15^2 (1 - 0.15)^8  \cdot 0.23}
         {\frac{1}{0.423} \, \phi \left ( \frac{-1.04 +  0.793}{0.423}\right )}\right) 
          =  0.080. \;
\end{align*}

For $\hat p_4(k = 2 \mid n = 10)^{(t+1)}$, we then get: 

\begin{align*}
 \label{E:BSRunEx}
          \hat p_4&(k = 2  \mid n = 10)^{(t+1)}   = \cfrac{\cfrac{1}{N_2} \displaystyle \sum_{i = 1}^{N_2} 
                      \cfrac{l_{2,i}}{s_1 l_{2,i} + s_2 \hat p_4(k = 2 \mid n = 10)^{(t)}}}
                     {\cfrac{1}{N_1} \displaystyle \sum_{j = 1}^{N_1}
                      \cfrac{1}{s_1 l_{1,j} + s_2 \hat p_4(k = 2 \mid n = 10)^{(t)}}} \\
\\
              & = \cfrac{\cfrac{1}{12} 
                      \left(\cfrac{0.077}{0.5 \cdot 0.077 + 0.5 \cdot \hat p_4(k = 2 \mid n = 10)^{(t)}}
            + \ldots +      \cfrac{ 0.084}{0.5 \cdot  0.084 + 0.5 \cdot \hat p_4(k = 2 \mid n = 10)^{(t)}} \right)}
                  {\cfrac{1}{12} 
                      \left(\cfrac{1}{0.5 \cdot  0.080 + 0.5 \cdot \hat p_4(k = 2 \mid n = 10)^{(t)}}
            + \ldots +      \cfrac{1}{0.5 \cdot  0.103 + 0.5 \cdot \hat p_4(k = 2 \mid n = 10)^{(t)}} \right)} \; .
 \end{align*}
 
Using $\hat p(y)^{(0)} = 0$, we obtain as updated estimate of the marginal likelihood $\hat p_4(k = 2  \mid n = 10)^{(1)} = 0.0908$. This iterative procedure has to be repeated until our predefined tolerance criterion is reached. For our running example, this criterion is reached after five iterations. We now obtain the bridge sampling estimate of the marginal likelihood as $\hat p_4(k = 2  \mid n = 10)^{(5)} = 0.0902$.

\subsection{Interim Summary}
So far we used the beta-binomial model to illustrate the computation of four different estimators of the marginal likelihood. These four estimators were discussed in order of increasing sophistication, such that the first three estimators provided the proper context for understanding the fourth, most general estimator---the bridge sampler. This estimator is the focus in the remainder of this tutorial. The goal of the next sections is to demonstrate that bridge sampling is particularly suitable to estimate the marginal likelihood of popular models in mathematical psychology. Importantly, bridge sampling may be used to obtain accurate estimates of the marginal likelihood of hierarchical models (for a detailed comparison of bridge sampling versus its special cases see \citeNP{fruhwirth2004estimating, sinharay2005empirical}).

\subsection{Assessing the Accuracy of the Bridge Sampling Estimate}

In this section we show how to quantify the accuracy of the bridge sampling estimate. A straightforward approach would be to apply the bridge sampling procedure multiple times and investigate the variability of the marginal likelihood estimate. In practice, however, this solution is often impractical due to the substantial computational burden of obtaining the posterior samples and evaluating the relevant quantities in the bridge sampling procedure.

\citeA{fruhwirth2004estimating} proposed an alternative approach that approximates the estimator's expected relative mean-squared error:
\begin{equation}
\label{E:RE2}
RE^2 = \frac{ \mathbb{E}\left [\big(\hat p_4(y) - p(y)\big)^2 \right ]}{p(y)^2}.
\end{equation}
The derivation of this approximate relative mean-squared error by \citeauthor{fruhwirth2004estimating} takes into account that the samples from the proposal distribution $g(\theta)$ are independent, whereas the MCMC samples from the posterior distribution $p(\theta | y)$ may be autocorrelated. The approximate relative mean-squared error is given by:
\begin{equation}
\label{E:RE2bridge}
\widehat{RE}^2 = \frac{1}{N_2} \frac{V_{g(\theta)}\big(f_1(\theta)\big)}{\mathbb{E}^2_{g(\theta)}\big(f_1(\theta)\big)} + \frac{\rho_{f_2}(0)}{N_1} \frac{V_{\text{post}}\big(f_2(\theta)\big)}{\mathbb{E}^2_{\text{post}}\big(f_2(\theta)\big)},
\end{equation}

\begin{sloppypar}
\noindent
where $f_1(\theta) = \frac{p(\theta \mid y)}{s_1 p(\theta \mid y) + s_2 g(\theta)}$, $f_2(\theta) = \frac{g(\theta)}{s_1 p(\theta \mid y) + s_2 g(\theta)}$, $V_{g(\theta)}\big(f_1(\theta)\big) = \int \left(f_1(\theta) - \mathbb{E}\left[f_1(\theta)\right]\right)^2 g(\theta) \thinspace \text{d}\theta$ denotes the variance of $f_1(\theta)$ with respect to the proposal distribution $g(\theta)$ (the variance $V_{\text{post}}\big(f_2(\theta)\big)$ is defined analogously), and $\rho_{f_2}(0)$ corresponds to the normalized spectral density of the autocorrelated process $f_2(\theta)$ at the frequency 0.
\end{sloppypar}

In practice, we approximate the unknown variances and expected values by the corresponding sample variances and means. Hence, for evaluating the variance and expected value with respect to $g(\theta)$, we use the $N_2$ samples for $\tilde \theta_i$ from the proposal distribution.
To evaluate the variance and expected value with respect to the posterior distribution, we use the second batch of $N_1$ samples $\theta^*_j$ from the posterior distribution which we also use in the iterative scheme for computing the marginal likelihood. Because the posterior samples are obtained via an MCMC procedure and are hence autocorrelated, the second term in Equation~\ref{E:RE2bridge} is adjusted by the normalized spectral density (for details see \citeNP{fruhwirth2004estimating}).\footnote{We estimate the spectral density at frequency zero by fitting an autoregressive model using the \texttt{spectrum0.ar()} function as implemented in the \texttt{coda} R package \cite{PlummerEtAl2006}.} To evaluate the normalized posterior density which appears in the numerator of $f_1(\theta)$ and the denominator of both $f_1(\theta)$ and $f_2(\theta)$, we use the bridge sampling estimate as normalizing constant. 

Note that, under the assumption that the bridge sampling estimator $\hat p_4(y)$ is an unbiased estimator of the marginal likelihood $p(y)$, the square root of the expected relative mean-squared error (Equation~\ref{E:RE2}) can be interpreted as the coefficient of variation (i.e., the ratio of the standard deviation and the mean; \citeNP{brown1998coefficient}). In the remainder of this article, we report the coefficient of variation to quantify the accuracy of the bridge sampling estimate.

\section{Case Study: Bridge Sampling for Reinforcement Learning Models}
In this section, we illustrate the computation of the marginal likelihood using bridge sampling in the context of a published data set \cite{Busemeyer2002253} featuring the Expectancy Valence (EV) model---a popular reinforcement learning (RL) model for the Iowa gambling task \cite<IGT;>{Bechara19947}. We first introduce the task and the model, and then use bridge sampling to estimate the marginal likelihood of the EV model implemented in both an individual-level and a hierarchical Bayesian framework. For the individual-level framework, we compare estimates obtained from bridge sampling to importance sampling estimates published in \citeA{SteingroeverIS}. For the hierarchical framework, we compare our results to estimates from the Savage-Dickey density ratio test \cite{dickey1971weighted, dickey1970weighted, wagenmakers2010SD, Wetzels20102094}.

\subsection{The Iowa Gambling Task}
In this section we describe the IGT (see also \citeNP{SteingroeverIntuitive, Steingroever2012, SteingroeverPSP2013, Steingroever2013Frontiers, SteingroeverFitSubm, SteingroeverIS}). Originally, \citeA{Bechara19947} developed the IGT to distinguish decision-making strategies of patients with lesions to the ventromedial prefrontal cortex from the ones of healthy controls (see also \citeNP{Bechara199818, Bechara199919, Bechara2000123}). During the last decades, the scope of application of the IGT has increased tremendously covering clinical populations with, for example, pathological gambling \cite{CavediniKel2002}, obsessive-compulsive disorder \cite{CavediniAnn2002}, psychopathic tendencies \cite{Blair200149}, and schizophrenia \cite{Bark2005131, Martino2007121}. 

The IGT is a card game that requires participants to choose, over several rounds, cards from four different decks in order to maximize their long-term net outcome \cite{Bechara19947, Bechara28021997}. The four decks differ in their payoffs, and two of them result in negative long-term outcomes (i.e., the bad decks), whereas the remaining two decks result in positive long-term outcomes (i.e., the good decks). After each choice, participants receive feedback on the rewards and losses (if any) associated with that card, as well as their running tally of net outcomes over all trials so far. Unbeknownst to the participants, the task (typically) contains 100 trials. 

\begin{table}[bt]
\small
\centering
\caption{\textit{Summary of the Payoff Scheme of the Traditional IGT as Developed by \protect\citeA{Bechara19947}}}
\begin{tabular}{lrrrr} % l = left aligned, r = right aligned
\hline % for horizontal lines
& Deck A & Deck B & Deck C & Deck D \\
\cline{2-5}
&Bad deck & Bad deck & Good deck & Good deck \\
&with fre- & with infre- & with fre- & with infre-\\
&quent losses & quent losses & quent losses & quent losses \\
\hline
Reward/trial & 100 & 100 & 50 & 50 \\
Number of losses/10 cards & 5 & 1 & 5 & 1 \\
Loss/10 cards & $-$1250 & $-$1250 & $-$250 & $-$250 \\
Net outcome/10 cards & $-$250 & $-$250 & 250 & 250\\
\hline
\label{T:IGT} % for text reference to this table
\end{tabular}
\end{table}
 
A crucial aspect of the IGT is whether and to what extent participants eventually learn to prefer the good decks because only choosing from the good decks maximizes their long-term net outcome. The good decks are typically labeled as decks C and D, whereas the bad decks are labeled as decks A and B. Table~\ref{T:IGT} presents a summary of the traditional payoff scheme as developed by \citeA{Bechara19947}. This table illustrates that decks A and B yield high constant rewards, but even higher unpredictable losses: hence, the long-term net outcome is negative. Decks C and D, on the other hand, yield low constant rewards, but even lower unpredictable losses: hence, the long-term net outcome is positive. In addition to the different payoff magnitudes, the decks also differ in the frequency of losses: decks A and C yield frequent losses, while decks B and D yield infrequent losses.

\subsection{The Expectancy Valence Model}

In this section, we describe the EV model (see also \citeNP{SteingroeverPSP2013, SteingroeverFitSubm, SteingroeverIS, SteingroeverIntuitive}). Originally proposed by \citeA{Busemeyer2002253}, the EV model is arguably the most popular model for the IGT (for references see~\citeNP{SteingroeverPSP2013}, and for alternative IGT models see~\citeNP{Ahn200832, dai2015improved, SteingroeverFitSubm, worthy2013decomposing, worthy2014comparison}). The model formalizes participants' performance on the IGT through the interaction of three model parameters that represent distinct psychological processes. The first model assumption is that after choosing a card from deck \emph{k}, \emph{k} $\in \{ 1, 2, 3, 4 \}$, on trial \emph{t}, participants compute a weighted mean of the experienced reward \emph{W(t)} and loss \emph{L(t)} to obtain the utility of deck \emph{k} on trial \emph{t}, \emph{v$_{k}(t)$}:

\begin{equation*}
  \displaystyle v_k(t) = (1 - w)W(t) + wL(t).
\end{equation*}

\noindent
The weight that participants assign to losses relative to rewards is the attention weight parameter \emph{w}. A small value of \emph{w}, that is, $w < .5$, is characteristic for decision makers who put more weight on the immediate rewards and can thus be described as reward-seeking, whereas a large value of \emph{w}, that is, $w > .5$, is characteristic for decision makers who put more weight on the immediate losses and can thus be described as loss-averse \cite{Ahn200832, Busemeyer2002253}. 

The EV model further assumes that decision makers use the utility of deck \emph{k} on trial \emph{t}, $v_{k}(t)$, to update only the expected utility of deck \emph{k}, $Ev_k(t)$; the expected utilities of the unchosen decks are left unchanged. This updating process is described by the Delta learning rule, also known as the Rescorla-Wagner rule \cite{Rescorla1972}:

\begin{equation*}
  \displaystyle Ev_k(t) = Ev_k(t-1) + a(v_k(t) - Ev_k(t-1)).
\end{equation*}

\noindent
If the experienced utility $v_{k}(t)$ is higher than expected, the expected utility of deck \emph{k} is adjusted upward. If the experienced utility  $v_k(t)$ is lower than expected, the expected utility of deck \emph{k} is adjusted downward. This updating process is influenced by the second model parameter---the updating parameter \emph{a}. This parameter quantifies the memory for rewards and losses. A value of \emph{a} close to zero indicates slow forgetting and weak recency effects, whereas a value of \emph{a} close to one indicates rapid forgetting and strong recency effects. For all models, we initialized the expectancies of all decks to zero, $Ev_k(0) = 0$ ($k \in \{ 1, 2, 3, 4 \}$). This setting reflects neutral prior knowledge about the payoffs of the decks.

In the next step, the model assumes that the expected utilities of each deck guide participants' choices on the next trial $t+1$. This assumption is formalized by the softmax choice rule, also known as the ratio-of-strength choice rule \cite{Luce1959}:

\begin{equation*}
  \displaystyle Pr[S_k(t + 1)] = \frac{e^{\theta(t) \cdot Ev_k(t)}}{\sum^4_{j=1}e^{\theta(t) \cdot Ev_j(t)}}.
  \label{E:Luce}
\end{equation*}

\noindent
The EV model uses this rule to compute the probability of choosing each deck on each trial. This rule contains a sensitivity parameter $\theta$ that indexes the extent to which trial-by-trial choices match the expected deck utilities. Values of $\theta$ close to zero indicate random choice behavior (i.e., strong exploration), whereas large values of $\theta$ indicate choice behavior that is strongly determined by the expected utilities (i.e., strong exploitation). 

\noindent
The EV model uses a trial-dependent sensitivity parameter $\theta(t)$, which also depends on the final model parameter, response consistency $c^\prime \in [-5, 5]$: 

\begin{equation*}
  \displaystyle \theta(t) = (t/10)^{c^\prime}.
  \label{E:Theta}  
\end{equation*}

\noindent
If $c^\prime$ is positive, successive choices become less random and more determined by the expected deck utilities; if $c^\prime$ is negative, successive choices become more random and less determined by the expected deck utilities, a pattern that is clearly non-optimal. We restricted the consistency parameter of the EV model to the range $[-2, 2]$ instead of the proposed range $[-5, 5]$ \cite{Busemeyer2002253}. This modification improved the estimation of the EV model and prevented the choice rule from producing numbers that exceed machine precision (see also \citeNP{SteingroeverFitSubm}).

In sum, the EV model has three parameters: (1) the attention weight parameter $w \in [0, 1]$, which quantifies the weight of losses over rewards; (2) the updating parameter $a \in [0, 1]$, which determines the memory for past expectancies; and (3) the response consistency parameter $c^\prime \in [-2, 2]$, which determines the balance between exploitation and exploration.

\subsection{Data}

We applied bridge sampling to a data set published by \citeA{Busemeyer2002253}. The data set consists of 30 healthy participants each contributing $T = 100$ IGT card selections (see \citeauthor{Busemeyer2002253} for more details on the data sets).\footnote{Note that we excluded three participants due to incomplete choice data.}

\subsection{Application of Bridge Sampling to an Individual-Level Implementation of the EV Model}

In this section we describe how we use bridge sampling to estimate the marginal likelihood of an individual-level implementation of the EV model. This implementation estimates model parameters for each participant separately. Accordingly, we also obtain a marginal likelihood of the EV model for every participant.

\subsubsection{Schematic execution of the bridge sampler}
To obtain the bridge sampling estimate of the marginal likelihood for each participant, we follow the steps outlined in Figure~\ref{F:algorithm}.

For each participant $s$, $s \in \{1, 2, \ldots, 30\}$, we proceed as follows:

\begin{enumerate}
\item \emph{For each parameter, we draw $2 N_1$ samples from the posterior distribution.}
 \\
Since \citeA{SteingroeverIS} already fit an individual-level implementation of the EV model separately to the data of each participant in \citeA{Busemeyer2002253}, we reuse their posterior samples (see \citeNP{SteingroeverIS}, for details on the prior distributions and model implementation). Note that they parameterized the model not in terms of $c^\prime \in [-2, 2]$, but in terms of $c = (c^\prime + 2)/4, \; c \in [0, 1]$, and in the remainder of this article, we also use this reparameterization.
\\
For each participant, we choose $2 N_1$ to match the number of samples obtained from \citeA{SteingroeverIS} which was at least $5,000$; however, whenever this number of samples was insufficient to ensure convergence of the Hamiltonian Monte Carlo (HMC) chains, \citeA{SteingroeverIS} repeated the fitting routine with $5,000$ additional samples. \citeA{SteingroeverIS} confirmed convergence of the HMC chains by reporting that all $\hat R$ statistics were below 1.05.
\item \emph{We choose a proposal distribution.} \\
We generalize our approach from the running example and use a multivariate normal distribution as a proposal distribution. 
\item \emph{We transform the first batch of $N_1$ posterior samples.} \\
Since we use a multivariate normal distribution as a proposal distribution, we have to transform all posterior samples to the real line using the probit transformation, that is, $\omega_{s,j}^{*} = \Phi^{-1}(w^{*}_{s,j})$, $\alpha_{s,j}^{*} = \Phi^{-1}(a^{*}_{s,j})$, $\gamma_{s,j}^{*} = \Phi^{-1}(c^{*}_{s,j})$, $j=\{1, 2, \ldots, N_1\}$.
\item \emph{We fit the proposal distribution to the first batch of $N_1$ probit-transformed posterior samples.} \\
We use method of moment estimates for the mean vector and the covariance matrix obtained from the first batch of $N_1$ probit-transformed posterior samples to specify our multivariate normal proposal distribution.
\item \emph{We draw $N_2$ samples from the proposal distribution.}\\
We use the R software to randomly draw $N_2$ samples from the proposal distribution obtained in step 4. We obtain $(\tilde{\omega}_{s,i}, \tilde{\alpha}_{s,i}, \tilde{\gamma}_{s,i})$ with $i \in \{1, 2, \ldots, N_2\}$.
\item \emph{We calculate $l_{2,i}$ for all $N_2$ samples from the proposal distribution.}\\
This step involves assessing the value of the unnormalized posterior and the proposal distribution for all $N_2$ samples from the proposal distribution. 
Before we can assess the value of the unnormalized posterior (i.e., the product of the likelihood and the prior), we have to derive how our transformation in step 3 affects the unnormalized posterior. \\
First, we derive how our transformation affects the likelihood. To evaluate the likelihood, we need to transform the probit-transformed samples back to the original parameter scales. That is, we evaluate the likelihood for $(\Phi(\tilde{\omega}_{s,i}), \Phi(\tilde{\alpha}_{s,i}), \Phi(\tilde{\gamma}_{s,i}))$. Before formalizing the likelihood of the observed choices of participant $s$, we define the following variables: We define $Ch_s(t)$ as a vector containing the sequence of choices made by participant $s$ up to and including trial $t$, and $X_s(t)$ as a vector containing the corresponding sequence of net outcomes. We now obtain the following expression for the likelihood of the observed choices of participant $s$:

\begin{align} \label{Eq:ML0}
        \begin{split}
& p(Ch_s(T) \mid \Phi(\tilde{\omega}_{s,i}), \Phi(\tilde{\alpha}_{s,i}), \Phi(\tilde{\gamma}_{s,i}), X_s(T - 1)) = \\ 
 & \; \; \; \; \; \; \; \; \; \; \; \; \; \; \; \; \; \; \; \; \; \; \; \; \; \; \; \prod_{t = 0}^{T - 1} \prod_{k = 1}^4 
                 Pr[S_k(t + 1)]
                \cdot \delta_k(t + 1).
        \end{split}
\end{align}

\noindent
Here $T$ is the total number of trials, $Pr[S_k(t + 1)]$ is the probability of choosing deck $k$ on trial $t+1$, and $\delta_k(t+1)$ is a dummy variable which is 1 if deck $k$ is chosen on trial $t+1$ and 0 otherwise. \\
Second, we have to derive how our transformation affects the priors on each EV model parameter to yield priors on the probit-transformed model parameters. Since \citeA{SteingroeverIS} used independent uniform priors on $[0,1]$ we obtain standard normal priors on the probit-transformed model parameters (see beta-binomial example and Appendix D for an explanation).
\item \emph{We transform the second batch of $N_1$ posterior samples.} \\
This is analogous to step 2.
\item{ \emph{We calculate $l_{1,j}$ for the second batch of $N_1$ probit-transformed samples from the posterior distribution.}}\\
This is analogous to step 6.
\item \emph{We run the iterative scheme (Equation~\ref{eq:f9}) until our predefined tolerance criterion is reached.}\\
We use a tolerance criterion and initialization analogous to the running example. Once convergence is reached, we receive an estimate of the marginal likelihood for each participant, and derive the coefficient of variation for each participant using Equation~\ref{E:RE2bridge}. The largest coefficient of variation is $2.07\%$ suggesting that the bridge sampler has low variance.\footnote{Note that this measure relates to the marginal likelihoods, not to the log marginal likelihoods.} 
\end{enumerate} 

\subsubsection{Assessing the accuracy of our implementation}
\begin{figure}[!bt]
	\centering
    \includegraphics[width=0.8\textwidth]{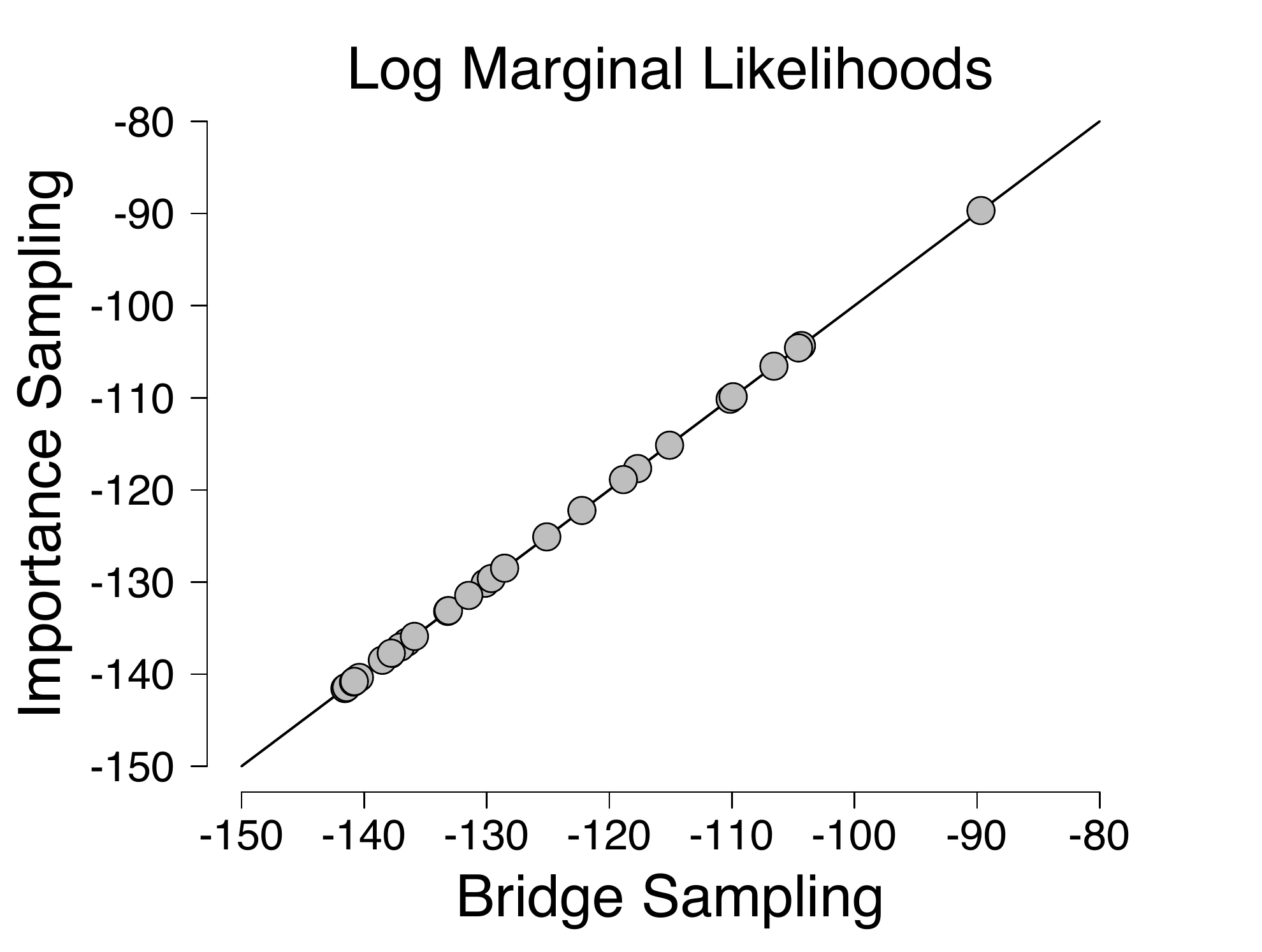}
	\caption{Comparison of the log marginal likelihoods obtained with bridge sampling (x-axis) and importance sampling reported by \protect\citeA{SteingroeverIS}(y-axis). The main diagonal indicates perfect correspondence between the two methods. Available at \url{https://tinyurl.com/yac3o8qs} under CC license \url{https://creativecommons.org/licenses/by/2.0/}.}
	\label{fig:nonehierarchical}
\end{figure}

To assess the accuracy of our implementation, we compared the marginal likelihood estimates obtained with our bridge sampler to the estimates obtained with importance sampling \cite{SteingroeverIS}. Figure \ref{fig:nonehierarchical} shows the log marginal likelihoods for the 30 participants of \citeA{Busemeyer2002253} obtained with bridge sampling (x-axis) and importance sampling reported by \citeauthor{SteingroeverIS} (2016; y-axis). The two sets of estimates correspond almost perfectly. These results indicate a successful implementation of the bridge sampler. Thus, this section emphasizes that both the importance sampler and bridge sampler can be used to estimate the marginal likelihood for the data of individual participants. However, when we want to estimate the marginal likelihood of a Bayesian hierarchical model, it may be difficult to find a suitable importance density. The bridge sampler, on the other hand, can be applied more easily and more efficiently.

\subsection{Application of Bridge Sampling to a Hierarchical Implementation of the EV Model}

In this section we illustrate how bridge sampling can be used to estimate the marginal likelihood of a hierarchical EV model.  This hierarchical implementation assumes that the parameters $w$, $a$, and $c$ from each participant are drawn from three separate group-level distributions. This model specification hence incorporates both the differences and the similarities between participants. We illustrate this application using again the \citeA{Busemeyer2002253} data set, and assume that these participants constitute one group.

\subsubsection{Schematic execution of the bridge sampler}

To compute the marginal likelihood, we again follow the steps outlined in Figure~\ref{F:algorithm}, with a few minor modifications.

\begin{enumerate}
\item \emph{For each parameter, that is, all individual-level and group-level parameters, we draw $2 N_1 = 60,000$ samples from the posterior distribution.} \\
To obtain the posterior samples, we fit a hierarchical Bayesian implementation of the EV model to the \citeA{Busemeyer2002253} data set using the software JAGS \cite{Plummer2003}.\footnote{We used a model file that is an adapted version of the model file used by \citeA{Ahn20114}.} We assume that, for each participant $s$, $s \in \{1, 2, \ldots, 30\}$, each probit-transformed individual-level parameter (i.e., $\omega_{s} = \Phi^{-1}(w_{s})$, $\alpha_{s} = \Phi^{-1}(a_{s})$, $\gamma_{s} = \Phi^{-1}(c_{s})$) is drawn from a group-level normal distribution characterized by a group-level mean and standard deviation parameter. For all group-level mean parameters $\mu_{\omega}, \mu_{\alpha}, \mu_{\gamma}$ we assume a standard normal distribution, and for all group-level standard deviation parameters $\sigma_{\omega}, \sigma_{\alpha}, \sigma_{\gamma}$ a uniform distribution ranging from $0$ to $1.5$. For a detailed explanation of the hierarchical implementation of the EV model, see \citeA{Wetzels201014}. \\
To reach convergence and reduce autocorrelation, we collect two MCMC chains, each with $120,000$ samples from the posterior distributions after having excluded the first $30,000$ samples as burn-in. Out of these $120,000$ samples per chain, we retained every fourth value yielding $30,000$ samples per chain. This setting resulted in all $\hat R$ statistics below 1.05 suggesting that all chains have successfully converged from their starting values to their stationary distributions. 
\item \emph{We choose a proposal distribution.}\\
We use a multivariate normal distribution as a proposal distribution. 
\item \emph{We transform the first batch of $N_1$ posterior samples.} \\
As before, we ensure that the range of the posterior distribution matches the range of the proposal distribution by using the probit transformation, that is, $\omega_{s,j}^{*} = \Phi^{-1}(w^*_{s,j})$, $\alpha_{s,j}^{*} = \Phi^{-1}(a^*_{s,j})$, $\gamma_{s,j}^{*} = \Phi^{-1}(c^*_{s,j})$, $\tau^{*}_{\omega, j} = \Phi^{-1}(({\sigma^{*}_{\omega, j}}) \, / \, {1.5} )$, $\tau^{ *}_{\alpha, j} = \Phi^{-1}(({\sigma^{*}_{\alpha, j}}) \, / \, {1.5} )$, and $\tau^{*}_{\gamma, j} = \Phi^{-1}(({\sigma^{*}_{\gamma, j}}) \, / \, {1.5} )$, $j=\{1, 2, \ldots, N_1\}$. The group-level mean parameters do not have to be transformed because they already range across the entire real line.
\item \emph{We fit the proposal distribution to the first batch of the $N_1$ probit-transformed posterior samples.}\\
We use method of moment estimates for the mean vector and the covariance matrix obtained from the first batch of $N_1$ probit-transformed posterior samples to specify our multivariate normal proposal distribution.
\item \emph{We draw $N_2$ samples from the proposal distribution.}\\
We use the R software to randomly draw $N_2$ samples from the proposal distribution obtained in step 4. We obtain $(\tilde{\omega}_{s,i}, \tilde{\alpha}_{s,i}, \tilde{\gamma}_{s,i})$ and $(\tilde \mu_{\omega, i}, \tilde \tau_{\omega, i}, \tilde \mu_{\alpha, i}, \tilde \tau_{\alpha, i}, \tilde \mu_{\gamma, i}, \tilde \tau_{\gamma, i})$ with $i \in \{1, 2, \ldots, N_2\}$ and $s \in \{1, 2, \ldots, 30\}$.
\item \emph{We calculate $l_{2,i}$ for all $N_2$ samples from the proposal distribution.} \\
This step involves assessing the value of the unnormalized posterior and the proposal distribution for all $N_2$ samples from the proposal distribution. The unnormalized posterior is defined as: \\
$\left( \prod_{s=1}^{30} p(Ch_s(T) \mid \Phi(\boldsymbol{\tilde \kappa_{s,i}}), X_s(T - 1)) \; p(\boldsymbol{\tilde \kappa_{s,i}} \mid \boldsymbol{\tilde \zeta}_i) \right)  \ p(\boldsymbol{\tilde \zeta}_i) $, where $Ch_s(T)$ refers to all choices of subject $s$, $X_s(T - 1)$ to the net outcomes that subject $s$ experienced on trials $1, 2, \ldots, T - 1$, $\boldsymbol{\tilde \kappa_{s,i}} = (\tilde \omega_{s,i}, \tilde \alpha_{s,i}, \tilde \gamma_{s,i})$ to the $i^{th}$ sample from the proposal distribution for the individual-level parameters of subject $s$, and $\boldsymbol{\tilde\zeta}_i$ to the $i^{th}$ sample from the proposal distribution for all group-level parameters (e.g., $\boldsymbol{\tilde \zeta}_i = (\tilde \mu_{\omega,i}, \tilde \tau_{\omega,i}, \tilde \mu_{\alpha,i}, \tilde \tau_{\alpha,i}, \tilde \mu_{\gamma,i}, \tilde \tau_{\gamma,i})$).
\\
The likelihood function for a given participant is the same as in the individual case. However, for each participant we now have to add besides the prior on the individual-level parameters also the prior on the group-level parameters. The product of the likelihood and the priors gives the unnormalized posterior density (see Appendix E for more details).
\item{We follow steps 7 -- 9, as outlined for the bridge sampler of the individual-level implementation of the EV model.}
\end{enumerate} 

\subsubsection{Assessing the accuracy of our implementation}
To investigate the accuracy of our implementation, we compare Bayes factors obtained with bridge sampling to Bayes factors obtained from the Savage-Dickey density ratio test 
(\citeNP{dickey1970weighted, dickey1971weighted}; for a tutorial, see \citeNP{wagenmakers2010SD}). The Savage-Dickey density ratio is a simple method for computing Bayes factors for nested
models. We artificially create three nested models by taking the full EV model $\mathcal{M}_f$ in which all parameters are free to vary, and then restricting one of the three group-level mean parameters, that is, $\mu_\omega$, $\mu_\alpha$, or $\mu_\gamma$, to a predefined value. For these values we choose the intersection point of the prior and posterior distribution of each group-level mean parameter. To obtain these intersection points, we fit the full EV model and then use a nonparametric logspline density estimator \cite{Logspline}. The obtained values are presented in Table~\ref{table:results_hier}. Since we compare the full model to each restricted model, we obtain three Bayes factors.

According to the Savage-Dickey density ratio test, the Bayes factor for the full model versus a specific restricted model $\mathcal{M}_r$ can be obtained by dividing the height of the prior density at the predefined parameter value $\theta_0$ by the height of the posterior at the same location:

\begin{align}
  \label{Eq:SavageDickey}
\text{BF}_{\mathcal{M}_{f}, \mathcal{M}_{r}} 
= \cfrac{p(y\mid \mathcal{M}_f)}{p(y\mid \mathcal{M}_r)} 
= \cfrac{p(\theta = \theta_0\mid \mathcal{M}_f)}{p(\theta = \theta_0\mid y, \mathcal{M}_f)}.
\end{align}

\begin{figure}
	\centering
    \includegraphics[width=0.8\textwidth]{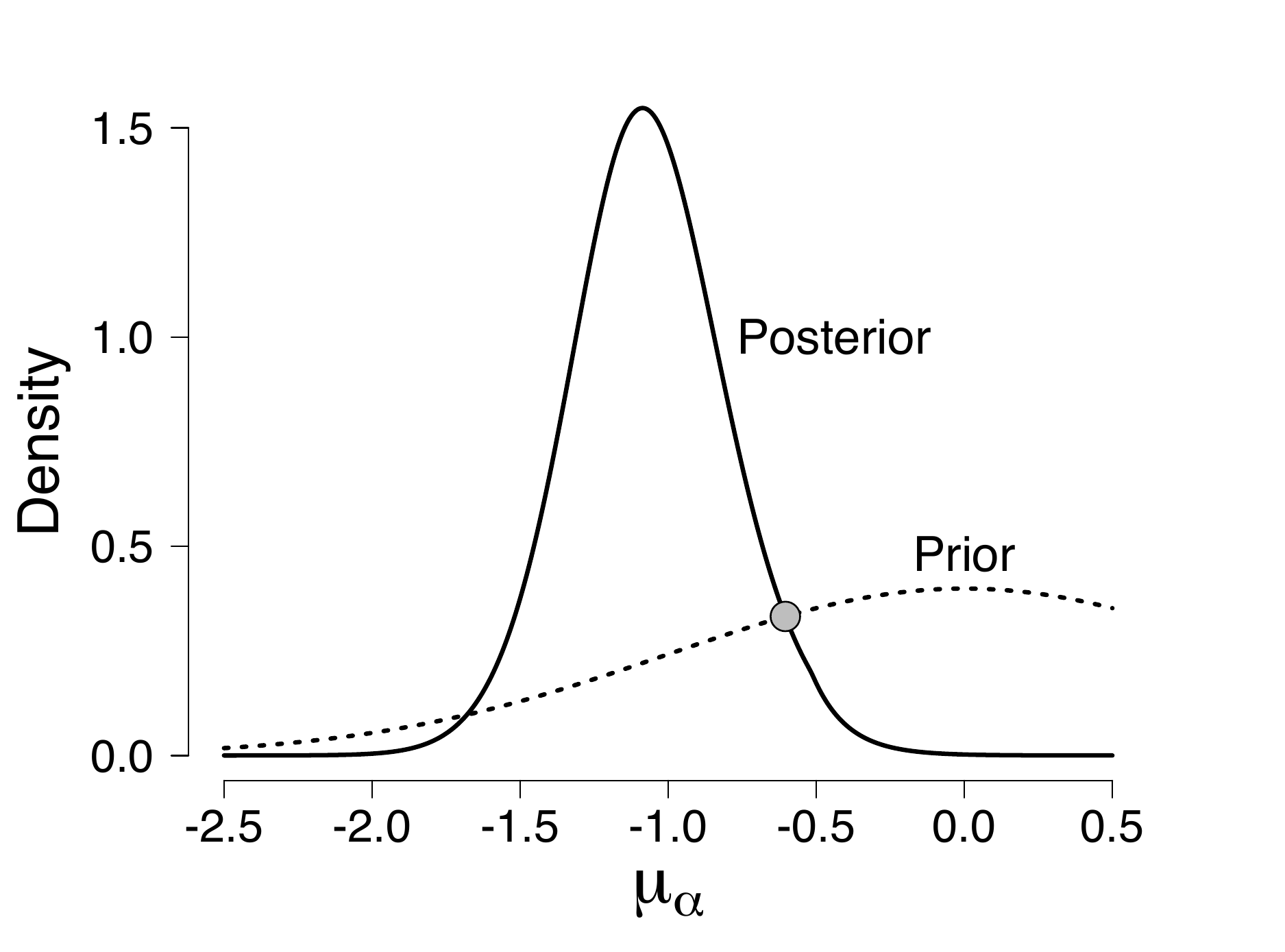}
	\caption{Prior and posterior distribution of the group-level
            mean $\mu_{\alpha}$ in the \protect\citeA{Busemeyer2002253} data set. The figure shows the posterior distribution (solid line) and the prior distribution (dotted line). The gray dot indicates the intersection of the prior and the posterior distributions, for which the Savage-Dickey Bayes factor equals $1$. Available at \url{https://tinyurl.com/y7cyxclq} under CC license \url{https://creativecommons.org/licenses/by/2.0/}.}
	\label{fig:savagedickey}
\end{figure}

Since we choose $\theta_0$ to be the intersection point of the prior and posterior distribution, $\text{BF}_{\mathcal{M}_{f}, \mathcal{M}_{r}}$ equals 1. This Savage-Dickey Bayes factor of 1 indicates that the marginal likelihood under the full model equals the marginal likelihood under the restricted model. Figure \ref{fig:savagedickey} illustrates the Savage-Dickey Bayes factor comparing the full model to the model assuming $\mu_{\alpha}$ fixed to $-0.604$. 

The computation of the three bridge sampling Bayes factors, on the other hand, works as follows: First, we follow the steps outlined above to obtain the bridge sampling estimate of the full EV model. Second, we obtain the bridge sampling estimate of the marginal likelihood for the three restricted models. This requires adapting the steps outlined above to each of the three restricted models. Lastly, we use the first equality in Equation~\ref{Eq:SavageDickey} to obtain the three Bayes factors.
 
The Bayes factors derived from bridge sampling are reported in Table \ref{table:results_hier}. It is evident that Bayes factors derived from bridge sampling closely approximate the Savage-Dickey Bayes factors of 1. These results suggest a successful implementation of the bridge sampler. This is also reflected by the close match between the log marginal likelihoods of the four models presented in the third column of Table \ref{table:results_hier}. 

Finally, we confirm that the bridge sampler has low variance; the coefficient of variation with respect to the marginal likelihood of the full model and the three restricted models ranges between $9.71$ and $16.44\%$.

\begin{table}
\caption {Bayes Factors Comparing the Full EV Model to the Restricted EV Models, Log Marginal Likelihoods, and Coefficient of Variation (With Respect to the Marginal Likelihood) Expressed as a Percentage}
\label{table:results_hier} 
\begin{center}
\begin{tabular}{lcp{2.3cm}r}
        \toprule
Model         & Bayes Factor & Log Marginal Likelihood & $CV [\%]$        \\\midrule 
full model                                          &   --      & $-3800.434$ & $10.13$  \\
restricted at $\mu_\omega = -0.334 $ &  $1.202$  & $-3800.618$ & $16.44$  \\
restricted at $\mu_\alpha = -0.604$ &  $1.052$  & $-3800.484$ & $9.71$  \\
restricted at $\mu_\gamma = 0.92 $  &  $1.068$  & $-3800.500$ & $12.03$  \\
\bottomrule
\end{tabular}
\end{center}
\end{table}

\section{Discussion}

In this tutorial, we explained how bridge sampling can be used to estimate the marginal likelihood of popular models in mathematical psychology. As a running example, we used the beta-binomial model to illustrate step-by-step the bridge sampling estimator. To facilitate the understanding of the bridge sampler, we first discussed three of its special cases---the naive Monte Carlo estimator, the importance sampling estimator, and the generalized harmonic mean estimator. Consequently, we introduced key concepts that became gradually more complicated and sophisticated. In the second part of this tutorial, we showed how bridge sampling can be used to estimate the marginal likelihood of both an individual-level and a hierarchical implementation of the Expectancy Valence (EV; \citeNP{Busemeyer2002253}) model---a popular reinforcement-learning model for the Iowa gambling task (IGT; \citeNP{Bechara19947}). The running example and the application of bridge sampling to the EV model demonstrated the positive aspects of the bridge sampling estimator, that is, its accuracy, reliability, practicality, and ease-of-implementation \cite{Dicicciokass1997, fruhwirth2004estimating, MengWong1996}.\par

The bridge sampling estimator is superior to the naive Monte Carlo estimator, the importance sampling estimator, and the generalized harmonic mean estimator for several reasons. First, \citeA{MengWong1996} showed that, among the four estimators discussed in this article, the bridge sampler presented in this article minimizes the mean-squared error because it uses the optimal bridge function. 
Second, in bridge sampling, choosing a suitable proposal distribution is much easier than choosing a suitable importance density for the importance sampling estimator or the generalized harmonic mean estimator because bridge sampling is more robust to the tail behavior of the proposal distribution relative to the posterior distribution. This advantage facilitates the application of the bridge sampler to higher-dimensional and complex models. This characteristic of the bridge sampler combined with the popularity of higher-dimensional and complex models in mathematical psychology suggests that bridge sampling can advance model comparison exercises in many areas of mathematical psychology (e.g., reinforcement-learning models, response time models, multinomial processing tree models, etc.).
Third, bridge sampling is relatively straightforward to implement. In particular, our step-by-step procedure can be easily applied to other models with only minor changes of the code (i.e., the unnormalized posterior and potentially the proposal function have to be adapted). In our opinion, this is one of the  most attractive features of bridge sampling: It is an accurate yet very generic method. Exploiting this generic characteristic, we have implemented the bridge sampling procedure in the \texttt{bridgesampling} R package \cite{bridgesampling} in order to maximize its accessibility.  \par 

Despite the numerous advantages of the bridge sampler, the take-home message of this tutorial is not that the bridge sampler should be used blindly. There exist a large variety of methods to approximate the marginal likelihood that differ in their efficiency.\footnote{In general, a large number of approaches for model selection exist which are based on MCMC posterior sampling and some of them are not based on approximating the models' marginal likelihoods \cite<e.g.,>{Ando2007,SpiegelhalterEtAl2002}. A comparison of these methods is beyond the scope of this tutorial.} The most appropriate method optimizes the trade-off between accuracy and implementation effort. This trade-off depends on a number of aspects such as the complexity of the model, the number of models under consideration, the statistical experience of the researcher, and the time available. This suggests that the choice of the method should be reconsidered each time a marginal likelihood needs to be obtained. Obviously, when the marginal likelihood can be determined analytically, bridge sampling is not needed at all. If the goal is to compare (at least) two nested models, the Savage-Dickey density ratio test \cite{dickey1970weighted, dickey1971weighted} might be a better alternative. Note, however, that this requires an approximation of the marginal posterior density of one or more parameters which can be unstable in case the test value falls in the tail of the distribution. If only an individual-level implementation of a model is used, importance sampling may be easier to implement and may require less computational effort. This presupposes that one can find a proposal distribution with fatter tails than the posterior which may not always be trivial (even in an individual-level case). If the goal is to obtain the marginal likelihood of a large number of relatively simple models, the product space or reversible jump method (RJMCMC) might be more appropriate \cite{carlin1995bayesian, green1995reversible,LodewyckxEtAl2011}. In contrast to bridge sampling, implementations of these methods tend to be problem-specific rather than generic (but see \citeNP{LunnEtAl2009generic}). If a researcher with a limited programming background and/or little time resources wants to conduct a model comparison exercise, rough approximations of the Bayes factor, such as the Bayesian information criterion, might be more suitable \cite{Schwarz1978}. It should be kept in mind, however, that this approximation assumes a certain prior structure that may not respect the knowledge or information that researchers have at their disposal. On the other hand, a researcher with an extensive background in programming and mathematical statistics might consider using path sampling---a generalization of bridge sampling \cite{GelmanMeng1998}.

To conclude, in this tutorial we showed that bridge sampling offers a reliable and easy-to-implement approach to estimating a model's marginal likelihood. Bridge sampling can be profitably applied to a wide range of problems in mathematical psychology involving parameter estimation, model comparison, and Bayesian model averaging.

\section*{Acknowledgements}

We thank \citeA{Busemeyer2002253} for providing the data used in this article. \\
Funding for this research was provided by the Berkeley Initiative for Transparency in the Social Sciences, a program of the Center for Effective Global Action (CEGA), with support from the Laura and John Arnold Foundation.
This work was supported in part by a Vici grant from the Netherlands Organisation for Scientific Research (NWO) to EJW (016.Vici.170.083).
This research was furthermore supported by an NWO grant to QFG (406-16-528), UB (406-12-125), and to HS (404-10-086), a European Research Council (ERC) grant to AL and EJW (283876), and a Veni grant (451-15-010) from the NWO to DM.

\bibliography{bibfile.bib}

\setcounter{table}{0}
\setcounter{figure}{0}
\setcounter{equation}{0}
\renewcommand\thefigure{0\arabic{figure}}
\renewcommand{\thetable}{0\arabic{table}}
\renewcommand{\theequation}{C\arabic{equation}}

\newpage
\appendix

\section{The Bridge Sampling Estimator as a General Case of Methods 1 -- 3}

In this section we show that the naive Monte Carlo, the importance sampling, and the generalized harmonic mean estimators are special cases of the bridge sampling estimator under specific choices of the bridge function $h(\theta)$ and the proposal distribution $g(\theta)$.\footnote{Note that bridge sampling is also a general case of the \citeA{Chib200l} method of estimating the marginal likelihood using the Metropolis-Hastings acceptance probability \cite{Meng2002, mira2004bridge}.} An overview is provided in Table~\ref{T:Summary}.

\begin{table}[h]
	\small
	\centering
	\caption{\textit{Summary of the Bridge Sampling Estimators for the Marginal Likelihood, and Its Special Cases: the Naive Monte Carlo, Importance Sampling, and Generalized Harmonic Mean Estimator}}
	\begin{tabular}{>{\raggedright}p{0.14\columnwidth}llll} % l = left aligned, r = right aligned
		\hline % for horizontal lines
		Method & Estimator & Samples & Bridge Function $h(\theta)$  \\	
		\hline	
		Bridge sampling & $\cfrac{\frac{1}{N_2} \sum_{i = 1}^{N_2} 
                             p(y\mid \tilde \theta_i)
                             \; p(\tilde \theta_i)
                             \; h(\tilde \theta_i)}
               {\frac{1}{N_1} \sum_{j = 1}^{N_1} 
                             h(\theta^*_j)
                             \; g(\theta^*_j)}$ &
                              $\tilde \theta_i \sim g(\theta)$ & $h(\theta) = \cfrac{C}{\frac{N_1}{N_2 + N_1} p(y\mid \theta)p(\theta) +
\frac{N_2}{N_2 + N_1} p(y)g(\theta)}$\\
 		&& $\theta^*_j \sim p(\theta\mid y)$ \\
\\        
		Naive Monte Carlo & $\cfrac{1}{N} \sum_{i = 1}^N p(y\mid \tilde \theta_i)$ &
        $\tilde \theta_i \sim p(\theta)$ & $h(\theta) = \cfrac{1}{g(\theta)}$ and $g(\theta) = p(\theta)$  \\
		\\
		Importance sampling & $\cfrac{1}{N} \mathlarger{\sum}_{i = 1}^N
        \cfrac{p(y\mid \tilde \theta_i) \; p(\tilde \theta_i)}
             {g_{IS}(\tilde \theta_i)}$  & $\tilde \theta_i \sim g_{IS}(\theta)$ & $h(\theta) = \cfrac{1}{g(\theta)}$ and $g(\theta) = g_{IS}(\theta)$ \\
		\\
		Generalized harmonic mean & $\left( \cfrac{1}{N} \mathlarger{\sum}_{i = 1}^N 
     \cfrac{g_{IS}(\theta^*_i)}{p(y\mid \theta^*_i)
                 \; p(\theta^*_i)}
        \right)^{-1}$ & $\theta^*_i \sim p(\theta\mid y)$ & $h(\theta) = \cfrac{1}{p(y \mid \theta) p(\theta)}$  and $g(\theta) = g_{IS}(\theta)$ \\
		\\
		\hline
	\multicolumn{4}{l}  {\footnotesize{\textit{Note.} $p(\theta)$ is the prior distribution, $g_{IS}(\theta)$ is the importance density, $p(\theta | y)$ is the posterior distribution, $g(\theta)$ is the}}\\ 
	 \multicolumn{4}{l}  {\footnotesize{proposal distribution, $h(\theta)$ is the bridge function, and $C$ is a constant. The last column shows the bridge function}}\\
	 \multicolumn{4}{l} {\footnotesize{needed to obtain the special cases.}}
	\label{T:Summary} % for text reference to this table	
	\end{tabular}		
\end{table}

To prove that the bridge sampling estimator reduces to the naive Monte Carlo estimator, consider bridge sampling, choose the prior distribution as the proposal distribution (i.e., $g(\theta) = p(\theta )$), and specify the bridge function as $h(\theta) = 1/{g(\theta)}$. Inserting these specifications into Equation~\ref{eq:f7} yields:

\begin{align*} 
  \begin{split}
       \hat p_4 \Big( y \mid h(\theta) = \cfrac{1}{g(\theta)}, \;  g(\theta) = p(\theta) \Big)
  &= \cfrac{\frac{1}{N_2} \sum_{i = 1}^{N_2} 
                              \cfrac{1}{p(\tilde \theta_i)}
                             \; p(y\mid \tilde \theta_i)
                             \; p(\tilde \theta_i)}
               {\frac{1}{N_1} \sum_{j = 1}^{N_1} 
                             \cfrac{1}{p(\theta^*_j)}
                             \; p(\theta^*_j)},  \ \ \tilde \theta_i \sim p(\theta), \ \ \theta^*_j \sim p(\theta\mid y) \\
                             \\
  &= \cfrac{\frac{1}{N_2} \sum_{i = 1}^{N_2} 
                             p(y\mid \tilde \theta_i)}
               {\frac{1}{N_1}  N_1} 
  = \frac{1}{N_2} \sum_{i = 1}^{N_2} 
                              p(y\mid \tilde \theta_i) \; , \ \ \tilde \theta_i \sim p(\theta), \;                                       
  \end{split}                             
\end{align*}

\noindent
which is equivalent to the naive Monte Carlo estimator shown in Equation~\ref{Eq:NMCE}.

To prove that the bridge sampling estimator reduces to the importance sampling estimator, consider bridge sampling, choose the importance density as the proposal distribution (i.e., $g(\theta) = g_{IS}(\theta)$), and specify the bridge function as $h(\theta) = {1} / {g(\theta)} \,$. Inserting these specifications into Equation~\ref{eq:f7} yields:

\begin{align*} 
  \begin{split}
       \hat p_4 \Big(y \mid h(\theta) = \cfrac{1}{g(\theta)}, \;  g(\theta) = g_{IS}(\theta)  \Big)
             &= \cfrac{\frac{1}{N_2} \sum_{i = 1}^{N_2} 
                              \cfrac{1}{g_{IS}(\tilde \theta_i)}
                             \; p(y\mid \tilde \theta_i)
                             \; p(\tilde \theta_i)}
               {\frac{1}{N_1} \sum_{j = 1}^{N_1} 
                             \cfrac{1}{g_{IS}(\theta^*_j)}
                             \; g_{IS}(\theta^*_j)}, \ \ \tilde \theta_i \sim g_{IS}(\theta), \ \ \theta^*_j \sim p(\theta\mid y) \\
                             \\
  &= \cfrac{\frac{1}{N_2} \sum_{i = 1}^{N_2} 
                              \cfrac{p(y\mid \tilde \theta_i)\;p(\tilde \theta_i) }{g_{IS}(\tilde \theta_i)}}
               {\frac{1}{N_1}  N_1}  
     = \frac{1}{N_2} \sum_{i = 1}^{N_2} 
                              \cfrac{p(y\mid \tilde \theta_i) \; p(\tilde \theta_i)}{g_{IS}(\tilde \theta_i)} \; , \ \ \tilde \theta_i \sim g_{IS}(\theta),                                        
  \end{split}                             
\end{align*}

\noindent
which is equivalent to the importance sampling estimator shown in Equation~\ref{Eq:ISE}.
   
% GHME
To prove that the bridge sampling estimator reduces to the generalized harmonic mean estimator, consider bridge sampling, choose the importance density as the proposal distribution (i.e., $g(\theta) = g_{IS}(\theta)$), and specify the bridge function as $h(\theta) = {1} / ({p(y\mid \theta) \; p(\theta)})$. Inserting these specifications into Equation~\ref{eq:f7} yields:

\begin{align*} 
  \begin{split}
      & \hat p_4 \Big(y \mid h(\theta) = \cfrac{1}{p(y\mid \theta) \; p(\theta)}, \;  g(\theta) = g_{IS}(\theta) \Big) \\
             &= \cfrac{\frac{1}{N_2} \sum_{i = 1}^{N_2} 
                              \cfrac{1}{p(y\mid \tilde \theta_i) \; p(\tilde \theta_i)}
                             \; p(y\mid \tilde \theta_i)
                             \;  p(\tilde \theta_i)}
               {\frac{1}{N_1} \sum_{j = 1}^{N_1} 
                             \cfrac{1}{p(y\mid \theta^*_j) \; p(\theta^*_j)}
                             \; g_{IS}(\theta^*_j)}, \ \ \tilde \theta_i \sim g_{IS}(\theta), \ \ \theta^*_j \sim p(\theta\mid y) \\
                             \\
 &= \cfrac{\frac{1}{N_2}N_2}
                {\frac{1}{N_1} \sum_{j = 1}^{N_1} 
                             \cfrac{g_{IS}(\theta^*_j)}{p(y\mid \theta^*_j) \; p(\theta^*_j)}} 
 = \left( \frac{1}{N_1} \sum_{j = 1}^{N_1} 
                             \cfrac{g_{IS}(\theta^*_j)}{p(y\mid \theta^*_j) \; p(\theta^*_j)} \right) ^{-1} , \ \theta^*_j \sim p(\theta\mid y),                                       
  \end{split}                             
\end{align*}

\noindent
which is equivalent to the generalized harmonic mean estimator shown in Equation~\ref{Eq:GHME}.

\section{Bridge Sampling Implementation: Avoiding Numerical Issues}

In order to avoid numerical issues, we can rewrite Equation~\ref{eq:f9} in the following way:
\begin{align*}
\hat p_4 (y)^{(t+1)} &= \frac{\frac{1}{N_2}\sum\limits_{i = 1}^{N_2}\frac{l_{2,i}}{s_1 \thinspace l_{2,i} + s_2 \thinspace \hat p_4 (y)^{(t)}}}{\frac{1}{N_1}\sum\limits_{j = 1}^{N_1}\frac{1}{s_1 \thinspace l_{1,j} + s_2 \thinspace \hat p_4 (y)^{(t)}}}\\
\\
&= \frac{\frac{1}{N_2}\sum\limits_{i=1}^{N_2}\frac{\exp\big(\log( l_{2,i})\big)}{s_1 \exp\big(\log( l_{2,i})\big) + s_2 \hat p_4 (y)^{(t)} } }{\frac{1}{N_1}\sum\limits_{j=1}^{N_1}\frac{1}{s_1 \exp\big(\log( l_{1,j})\big) + 	s_2 \hat p_4 (y)^{(t)} }}\\
\\
&= \frac{\frac{1}{N_2}\sum\limits_{i=1}^{N_2}\frac{\exp\big(\log( l_{2,i})\big) \exp\big(-l^\ast\big)}{s_1 \exp\big(\log( l_{2,i})\big)\exp\big(-l^\ast\big) + s_2 \hat p_4 (y)^{(t)}\exp\big(-l^\ast\big) } }{\frac{1}{N_1}\sum\limits_{j=1}^{N_1}\frac{\exp\big(-l^\ast\big)}{s_1 \exp\big(\log( l_{1,j})\big)\exp\big(-l^\ast\big) + 	s_2 \hat p_4 (y)^{(t)}\exp\big(-l^\ast\big) }}\\
\\
&= \frac{1}{\exp\big(-l^\ast\big)}\frac{\frac{1}{N_2}\sum\limits_{i=1}^{N_2}\frac{\exp\big(\log( l_{2,i})-l^\ast\big)}{s_1 \exp\big(\log( l_{2,i})-l^\ast\big) + s_2 \hat p_4 (y)^{(t)}\exp\big(-l^\ast\big) } }{\frac{1}{N_1}\sum\limits_{j=1}^{N_1}\frac{1}{s_1 \exp\big(\log( l_{1,j})-l^\ast\big) + 	s_2 \hat p_4 (y)^{(t)}\exp\big(-l^\ast\big) }}\\
\\
&= \exp\big(l^\ast\big)\frac{\frac{1}{N_2}\sum\limits_{i=1}^{N_2}\frac{\exp\big(\log( l_{2,i})-l^\ast\big)}{s_1 \exp\big(\log( l_{2,i})-l^\ast\big) + 	s_2 \hat p_4 (y)^{(t)}\exp\big(-l^\ast\big) } }{\frac{1}{N_1}\sum\limits_{j=1}^{N_1}\frac{1}{s_1 \exp\big(\log( l_{1,j})-l^\ast\big) + s_2 \hat p_4 (y)^{(t)}\exp\big(-l^\ast\big) }}.
\end{align*}
$l^\ast$ is a constant which we can choose in a way that keeps the terms in the sums manageable.
We used $l^\ast~=~\text{median}(\log(l_{1,j}))$.
Let
\begin{equation*}
\hat{r}^{(t)} = \hat p_4 (y)^{(t)}\exp\big(-l^\ast\big),
\end{equation*}
so that
\begin{equation*}
\hat p_4 (y)^{(t)} = \hat{r}^{(t)} \exp\big(l^\ast\big).
\end{equation*}
Then we obtain
\begin{equation*}
\begin{split}
\hat p_4 (y)^{(t+1)} &= \exp\big(l^\ast\big)\frac{\frac{1}{N_2}\sum\limits_{i=1}^{N_2}\frac{\exp\big(\log( l_{2,i})-l^\ast\big)}{s_1 \exp\big(\log( l_{2,i})-l^\ast\big) + 	s_2 \hat{r}^{(t)} } }{\frac{1}{N_1}\sum\limits_{j=1}^{N_1}\frac{1}{s_1 \exp\big(\log( l_{1,j})-l^\ast\big) + s_2 \hat{r}^{(t)} }}\\
\\
\hat p_4 (y)^{(t+1)}\exp\big(-l^\ast\big) &= \frac{\frac{1}{N_2}\sum\limits_{i=1}^{N_2}\frac{\exp\big(\log( l_{2,i})-l^\ast\big)}{s_1 \exp\big(\log( l_{2,i})-l^\ast\big) + 	s_2 \hat{r}^{(t)} } }{\frac{1}{N_1}\sum\limits_{j=1}^{N_1}\frac{1}{s_1 \exp\big(\log( l_{1,j})-l^\ast\big) + s_2 \hat{r}^{(t)} }}\\
\\
\hat{r}^{(t+1)} &= \frac{\frac{1}{N_2}\sum\limits_{i=1}^{N_2}\frac{\exp\big(\log( l_{2,i})-l^\ast\big)}{s_1 \exp\big(\log( l_{2,i})-l^\ast\big) + 	s_2 \hat{r}^{(t)} } }{\frac{1}{N_1}\sum\limits_{j=1}^{N_1}\frac{1}{s_1 \exp\big(\log( l_{1,j})-l^\ast\big) + s_2 \hat{r}^{(t)} }}.
\end{split}
\end{equation*}
Hence, we can run the iterative scheme with respect to $\hat{r}$ which is more convenient because it keeps the terms in the sums manageable and multiply the result by $\exp(l^\ast)$ to obtain the estimate of the marginal likelihood or, equivalently, we can take the logarithm of the result and add $l^\ast$ to obtain an estimate of the logarithm of the marginal likelihood. 

\section{Correcting for the Probit Transformation}

In this section we describe how the probit transformation affects our expression of the generalized harmonic mean estimator (Equation~\ref{Eq:GHME}) to yield Equation~\ref{Eq:GHMEprobit}. Recall that we derived the generalized harmonic mean estimator using the following equality:

\begin{equation}
  \label{Eq:DeriveGHME}
  \begin{split}
    \cfrac{1}{p(y)} 
    = \int \cfrac{g_{IS}(\theta) }{p(y\mid \theta) p(\theta)} \; p(\theta \mid y) \; \mathrm{d}\theta.
  \end{split}
\end{equation}

For practical reasons, in the running example, we used a normal distribution on $\xi$ as importance density. This $\xi$ was defined as the probit transform of $\theta$	 (i.e, $\xi = \Phi^{-1}(\theta)$). In particular, the normal importance density was given by $\frac{1}{\hat \sigma} \phi \left ( \frac{\xi - \hat \mu}{\hat \sigma}\right )$. Note that this importance density is a function of $\xi$, whereas the general importance density $g_{IS}$ in Equation~\ref{Eq:DeriveGHME} is specified in terms of $\theta$. Therefore, to include our specific importance density into Equation~\ref{Eq:DeriveGHME}, we need to write it in terms of $\theta$. This yields $\frac{1}{\hat \sigma} \phi \left ( \frac{\Phi^{-1}(\theta) - \hat \mu}{\hat \sigma}\right ) \frac{1}{\phi\left ( \Phi^{-1} (\theta) \right )}$, where the latter factor comes from applying the change-of-variable method. Replacing $g_{IS}(\theta)$ in Equation~\ref{Eq:DeriveGHME} by this expression, results in:

\begin{equation}
  \label{Eq:DeriveGHME2}
  \begin{split}
    \cfrac{1}{p(y)} 
    &= \mathlarger{\int} \cfrac{ \frac{1}{\hat \sigma} \phi \left ( \frac{\Phi^{-1}(\theta) - \hat \mu}{\hat \sigma}\right ) \frac{1}{\phi\left ( \Phi^{-1} (\theta) \right )} }{p(y\mid \theta) p(\theta)} \; p(\theta \mid y) \; \mathrm{d}\theta \\
\\
&= \mathbb{E}_\text{post}
                           \left(\cfrac{ \frac{1}{\hat \sigma} \phi \left ( \frac{\Phi^{-1}(\theta) - \hat \mu}{\hat \sigma}\right ) \frac{1}{\phi\left ( \Phi^{-1} (\theta) \right )} }{p(y\mid \theta) \;
                           p(\theta)}\right)  .    
  \end{split}
\end{equation}

Rewriting results in:

\begin{align*}
               p(y) &= \left( \mathbb{E}_\text{post}
                           \left(\cfrac{ \frac{1}{\hat \sigma} \phi \left ( \frac{\Phi^{-1}(\theta) - \hat \mu}{\hat \sigma}\right ) \frac{1}{\phi\left ( \Phi^{-1} (\theta) \right )} }{p(y\mid \theta) \;
                           p(\theta)}\right) \right) ^{-1},
\end{align*}

which can be approximated as:

\begin{align} \label{Eq:GHMExi}
\begin{split}
        \hat p_{3}(y) &= \left( \cfrac{1}{N} \sum_{j = 1}^N 
     \cfrac{\overbrace{\frac{1}{\hat \sigma} \phi \left ( \frac{\Phi^{-1}(\theta^*_j) - \hat \mu}{\hat \sigma}\right ) \frac{1}{\phi\left ( \Phi^{-1} (\theta^*_j) \right )}}^\text{importance density}}{\underbrace{p(y\mid \theta^*_j)}_\text{likelihood}
                 \; \underbrace{p(\theta^*_j)}_\text{prior}}
        \right) ^{-1}, \; \; \underbrace{\theta^*_j \sim p(\theta\mid y) \; .}_{\substack{\text{samples from the}\\ \text{posterior distribution}}} \\
\\
&= \left( \cfrac{1}{N} \sum_{j = 1}^N 
     \cfrac{\overbrace{\frac{1}{\hat \sigma} \phi \left ( \frac{\xi^*_j - \hat \mu}{\hat \sigma}\right ) }^\text{importance density}}{\underbrace{p \left ( y\mid \Phi \left ( \xi^*_j \right ) \right )}_\text{likelihood}
                 \; \underbrace{p \left ( \Phi \left ( \xi^*_j \right ) \right ) \phi\left ( \xi^*_j \right ) }_\text{prior}}
        \right) ^{-1}, \; \; \underbrace{\xi^*_j = \Phi^{-1}(\theta^*_j) \; \text{and} \; \theta^*_j \sim p(\theta\mid y) \; ,}_{\substack{\text{probit-transformed samples}\\ \text{from the posterior distribution}}}     
\end{split}        
 \end{align}

\noindent
which shows that the generalized harmonic estimate can be obtained using the samples from the posterior distribution, or the probit-transformed ones. In the online-provided code, we use the latter approach (see also \citeNP{overstall2010default}). Note the in our running example,  $\forall \xi^*_j: \; p\left ( \Phi \left ( \xi^*_j \right ) \right ) = 1$.
 
\section{Details on the Application of Bridge Sampling to the Individual-Level EV Model}

In this section, we provide more details on how we obtained the unnormalized posterior distribution for a specific participant $s$, $s \in \{1, 2, \ldots, 30\}$. Since we focus on one specific participant, we drop the subscript $s$ in the remainder of this section. As explained in Appendix B, we run the iterative scheme with respect to $\hat{r}$ to avoid numerical issues. Consequently, we have to compute $\log(l_{1,j})$ and $\log(l_{2,i})$. Using $\boldsymbol{\tilde \kappa_i} = (\tilde \omega_i, \tilde \alpha_i, \tilde \gamma_i)$ for the $i^{th}$ sample from the proposal distribution, we get for $\log(l_{2,i})$ ($\log(l_{1,j})$ works analogously): 

\begin{align*} 
 \log(l_{2,i}) = \log \left( \cfrac{p(Ch(T)\mid \Phi(\boldsymbol{\tilde \kappa}_i), X(T - 1)) \; p(\Phi(\boldsymbol{\tilde \kappa}_i)) \; \phi(\boldsymbol{\tilde \kappa_i})}{g(\boldsymbol{\tilde \kappa}_i)} \right ).
\end{align*}

Therefore, instead of computing the unnormalized posterior distribution directly, we compute the logarithm of the unnormalized posterior distribution:

\begin{align*} 
  \label{}
        \begin{split}
  \log(p(Ch(T) \mid \Phi(\boldsymbol{\tilde \kappa}_i), X(T - 1)) \; p(\Phi(\boldsymbol{\tilde \kappa}_i)) \; \phi(\boldsymbol{\tilde \kappa}_i) ) \; 
       &= \log(p(Ch(T) \mid \Phi(\boldsymbol{\tilde \kappa}_i), X(T - 1))) + \\ & \quad \log(\phi(\tilde \omega_i)) + \log(\phi(\tilde \alpha_i)) + \log(\phi(\tilde \gamma_i)) ,
        \end{split}
\end{align*}

\noindent
because we assumed independent priors on each model parameter $w, a, c$. $\log(p(\Phi(\boldsymbol{\tilde \kappa}_i))) = 0$ because $p$ refers to the uniform prior on $[0, 1]$. 

\section{Details on the Application of Bridge Sampling to the Hierarchical EV Model}

Analogous to the last section, we explain here how we obtained the logarithm of the unnormalized posterior for the hierarchical implementation of the EV model. Using $\boldsymbol{\tilde \kappa_{s,i}} = (\tilde \omega_{s,i}, \tilde \alpha_{s,i}, \tilde \gamma_{s,i})$ for the $i^{th}$ sample from the proposal distribution for the individual-level parameters of subject $s$, and $\boldsymbol{\tilde \zeta}_i$ for the $i^{th}$ sample from the proposal distribution for all group-level parameters (i.e., $\boldsymbol{\tilde \zeta}_i = (\tilde \mu_{\omega,i}, \tilde \tau_{\omega,i}, \tilde \mu_{\alpha,i}, \tilde \tau_{\alpha,i}, \tilde \mu_{\gamma,i}, \tilde \tau_{\gamma,i})$), we get:

\begin{align*}
& \log \left( \left( \prod_{s=1}^{30} p(Ch_s(T) \mid \Phi(\boldsymbol{\tilde \kappa_{s,i}}), X_s(T - 1)) \; p(\boldsymbol{\tilde \kappa_{s,i}} \mid \boldsymbol{\tilde \zeta}_i) \right)  \; p(\boldsymbol{\tilde \zeta}_i) \right) \\
\\
&= \sum_{s = 1}^{N}  \left [ \log (p(Ch_s(T) \mid \Phi(\boldsymbol{\tilde \kappa}_{s,i}), X_s(T - 1))) + \right. \notag \\
\\
 & \quad  \left. {} \log \left( \cfrac{1}{1.5 \Phi(\tilde \tau_{\omega,i})} \; \phi\left (\cfrac{\tilde \omega_{s,i} - \tilde \mu_{\omega,i}}{1.5 \Phi(\tilde \tau_{\omega,i})} \right) \right) +
\log \left( \cfrac{1}{1.5 \Phi(\tilde \tau_{\alpha,i})} \;  \phi\left (\cfrac{\tilde \alpha_{s,i} - \tilde \mu_{\alpha,i}}{1.5 \Phi(\tilde  \tau_{\alpha,i})} \right) \right) + \right. \notag \\
\\
 & \quad \left. {} \log \left( \cfrac{1}{1.5 \Phi(\tilde \tau_{\gamma,i})} \;  \phi\left (\cfrac{\tilde \gamma_{s,i} - \tilde \mu_{\gamma,i}}{1.5 \Phi(\tilde \tau_{\gamma,i})} \right) \right) 
 \right ] + \\
\\
& \quad  \log \left( \phi(\tilde \mu_{\omega, i}) \right) + \log \left( \phi(\tilde \mu_{\alpha, i}) \right) + \log \left( \phi(\tilde \mu_{\gamma, i}) \right) + \\
\\
& \quad \; \log \left( \phi(\tilde \tau_{\omega, i})  \right) + \log \left( \phi(\tilde \tau_{\alpha,i})  \right) + \log \left( \phi(\tilde \tau_{\gamma,i})  \right).
\end{align*}

\end{document}